\documentclass[sigconf]{acmart}
\usepackage{todonotes}
\usepackage{xspace}
\usepackage{subcaption}
\usepackage[shortlabels]{enumitem}
\renewcommand\footnotetextcopyrightpermission[1]{}
\AtBeginDocument{%
  }

\copyrightyear{2026}
\acmYear{2026}
\setcopyright{cc}
\setcctype{by}
\acmConference[SCA/HPCAsiaWS 2026]{SCA/HPCAsia 2026 Workshops: Supercomputing Asia and International Conference on High Performance Computing in Asia Pacific Region Workshops}{January 26--29, 2026}{Osaka, Japan}
\acmBooktitle{SCA/HPCAsia 2026 Workshops: Supercomputing Asia and International Conference on High Performance Computing in Asia Pacific Region Workshops (SCA/HPCAsiaWS 2026), January 26--29, 2026, Osaka, Japan}

\acmDOI{10.1145/3784828.3784830}
\acmISBN{979-8-4007-2328-5/2026/01}
\begin{document}

\title{Understanding LLM Checkpoint/Restore I/O Strategies and Patterns}

\author{Mikaila J. Gossman}
\affiliation{%
  \institution{Clemson University}
  \city{Clemson}
  \state{IL}
  \country{USA}}
\email{mikailg@g.clemson.edu}

\author{Avinash Maurya}
\affiliation{%
  \institution{Argonne National Lab}
  \city{Lemont}
  \state{IL}
  \country{USA}
}
\email{amaurya@anl.gov}

\author{Bogdan Nicolae}
\affiliation{%
 \institution{Argonne National Lab}
 \city{Lemont}
 \state{IL}
 \country{USA}}
\email{bnicolae@anl.gov}

\author{Jon C. Calhoun}
\affiliation{%
  \institution{Clemson University}
  \city{Clemson}
  \state{SC}
  \country{USA}}
\email{jonccal@clemson.edu}

\newcommand{\proj}{\emph{ToolNameTDB}\xspace}
\newcommand{\AM}[1]{\todo[inline] {\footnotesize AM: #1}}
\renewcommand{\shortauthors}{Gossman et al.}

\begin{abstract}
As large language models (LLMs) and foundation models (FMs) scale, checkpoint-restore has become critical pattern for both training and inferences. This is a difficult pattern: LLMs and FMs are distributed at scale using 3D parallelism (tensor, pipeline, data), resulting in a large number of processes, each holding a large number of tensors of different shapes and sizes, which need to be captured to stable storage (e.g. parallel file systems) frequently. As a consequence, checkpoint and restore becomes a big data problem, exhibiting volume, variery and velocity. A key challenge is the need to traverse an entire storage stack, from the GPUs, where the tensors live, through the host memory, local storage and finally external repositories such as parallel file systems, where the checkpoints need to be persisted. Despite efficient I/O techniques (e.g. asynchronous flushing/prefetching), the memory tiers of the storage stack exhibit orders of magnitdue difference in I/O performance, leading to I/O bottlenecks, especially under concurrency. Kernel-accelerated I/O libraries such as \texttt{liburing} aim to address some of these I/O bottlenecks compared with the aging POSIX I/O interface. However, their effectiveness for LLM and checkpoint and restore remains largely unexplored. To this end, we develop a series of microbenchmarks that study various trade-off when using liburing. Specifically, we evaluate how aggregation, alignment, and I/O coalescing interact under buffered and direct I/O modes. Results show that uncoalesced, small-buffer operations halve throughput relative to synthetic workloads, while aggregation restores bandwidth and reduces metadata overhead compared to the state-of-the-art LLM checkpointing engines, achieving up to $3.9\times$ higher write throughput than DataStates-LLM and $7.6\times$ higher than TorchSnapshot. These findings underscore the importance of adopting file system–aware aggregation and I/O coalescing strategies in LLM checkpointing to fully exploit modern storage and I/O backends.
\end{abstract}


\maketitle

\section{Introduction}
\paragraph{\textbf{Context.}}
Large Language Models~(LLMs) have evolved exponentially over the past half-decade and continue to demonstrate widespread adoption across a wide range of industrial, commercial, and scientific applications. LLMs have shown remarkable fluency, accuracy, and coherence for use cases such as knowledge distillation~\cite{gou2021knowledge}, text summarization, literature search~\cite{tilwani2024reasons}, and complex reasoning~\cite{bubeck2023sparks}. Encouraged by such promising potential of LLMs, the more general Foundation Models (FMs) are gaining rapid traction, which combine multi-modal domain-specific data/instructions, capturing cross correlations through self and cross attention mechanisms~\cite{jiang2024mixtral, liu2024deepseek}. These models, coupled with agentic capabilities, drive advanced use cases, such as scientific assistants~\cite{gottweis2025towards}, automated data analysis, experimental design~\cite{bhatt2024experimental}, hypotheses formulation~\cite{manning2024automated}, self-evaluation, and critiquing of results~\cite{cappello2025eairaestablishingmethodologyevaluating}.

Unsurprisingly, the advancement of LLMs and FMs necessitates a substantial amount of resources and data for training, fine-tuning, and serving. For instance, the BLOOM-176B~\cite{bloom} model was trained on 384 GPUs for 3.5 months, accounting for $\sim$1M GPU hours. LLM inferences, on the other hand, are even more resource-intensive, amounting to 65\% of energy consumption of the AI clusters, whereas training and fine-tuning amount only to 35\%~\cite{castro2024AIenergy}. It is increasingly common to train and serve large models composed of billions or even trillions of parameters. At the same time, enterprises and organizations develop, refine, and serve multiple variations of these models, each tailored for different use cases. This results in a large number of variations of \emph{LLM checkpoints}, which together amount to an order of terabytes of data.

\paragraph{\textbf{Motivation}}
Given the long-running, large-scale distributed nature (using ``3D'' parallelism: data, tensor, pipeline) of LLM training and fine-tuning across thousands of GPUs~\cite{bloom, Meta2025Llama4}, it becomes essential to capture and persist LLM checkpoints for a variety of use cases: fault-tolerance, suspend-resume for accommodating high-priority jobs and walltime limits, studying model evolution, debugging / recovering from unstable training trajectories (e.g., model spikes), etc. Such LLM checkpoints need to be captured with high frequency (``velocity''), sometimes as frequently as every iteration (e.g. to capture and analyze differences introduced by specific mini-batches).
Furthermore, each checkpoint is composed of heterogeneous tensors (``variety''): parameters of layers of different shapes and sizes, optimizer states, etc. These heterogeneous tensors, distributed over a large number of GPUs, are grouped together into shards and persisted into a large number of files, which can grow to massive sizes (``volume''). As a consequence, capturing checkpoints is a big data problem that exhibits all three \emph{Vs}.

Restoring checkpoints is a big data problem as well. LLM checkpoints are loaded from persistent repositories at high frequency (``velocity'') in various scenarios: to resume training or perform fine-tuning at high frequency by opportunistically utilizing faster and cheaper preemptible instances (HPC) or spot instances (Cloud and Hyperscalars)~\cite{duan2024parcae,wu2025rlboost,athlur2022varuna}; to serve inference requests that need a large number of different models, all of which don't fit into the GPU memory at the same time and therefore need to be swapped in and out of slower memory tiers as needed. The same ``variety'' and ``volume'' as in the case of training applies to restored LLM checkpoints as well.

For both checkpoint and restore, a key challenge is the need to traverse an entire storage stack, from the GPUs, where the tensors live, through the host memory, local storage and finally external repositories such as parallel file systems, where the checkpoints need to be persisted. This leads to significant I/O pressure on the memory and interconnect subsystem because the faster memory tiers (GPU HBM, host DRAM) have fast interconnects (e.g., PCIe) to transfer checkpoints, but are limited in capacity; whereas the slower storage tiers (local disk, remote storage) have large capacities but perform slower read/writes and are connected through comparatively low-bandwidth interconnects (e.g., Infiniband). Without efficient I/O techniques, checkpoint and restore become a bottleneck that slow down both training and inferences. Thus, it is important
to study how the specific I/O patterns, caused by velocity, variety, and volume of LLM checkpoints, especially under concurrency due to 3D parallelism, affect the I/O performance and scalability of checkpoint and restore on modern HPC systems tailored for AI workloads.

\paragraph{\textbf{Challenges.}}
For large-scale LLM training, runtimes such as DeepSpeed~\cite{rasleyDeepSpeedSystemOptimizations2020} flush model states in parallel across GPUs, producing hundreds to thousands of distinct files per checkpoint, each capturing a portion of model or optimizer state~\cite{Maurya_2024}. This highly parallel write pattern accelerates checkpointing and simplifies the design of the runtime and checkpoint engine. However, the proliferation of large quantities of files on shared persistent storage causes metadata contention, leading to suboptimal I/O throughput and complex data management~\cite{lustre}. Recent studies~\cite{CARNEIRO2023104744} show that metadata-related operations alone account for up to 60\% of all file system operations, underscoring the non-trivial overheads imposed by frequent small-file creation and management. This results in a mismatch between ``simplifying application-centric checkpointing'' and ``I/O and storage-centric data-management'', which remains largely unoptimized for LLM and AI workloads~\cite{veloc, Maurya_2024, lightning_async_checkpoint_io}.

Furthermore, modern file systems and users continue to rely on buffered I/O by default, an approach originally optimized for slow spinning disks and early generation SSDs. Current HPC PFSes and SSDs now sustain multi-GB/s throughput per node, but buffered reads and writes can introduce redundant memory copies between the compute node's DRAM and file system caches. These extra copies inflate access latencies and amplify contention on metadata and I/O servers.
These extra copies inflate access latencies and amplify contention on metadata and I/O servers. Traditional benchmarking tools such as IOR~\cite{Shan2007_IOR} capture these effects only coarsely, as they are limited to POSIX semantics and simplified access patterns (e.g., file-per-process or a single aggregated file). Moreover, IOR assumes uniform object and access sizes across processes, making it unable to represent the heterogeneous and fine-grained data distributions observed in LLM checkpointing, where model states and optimizer tensors vary substantially in size across ranks. These gaps highlight the need for newer benchmarking strategies capable of modeling LLM checkpoint workloads and supporting backends beyond POSIX.

Compounding these factors, LLM checkpointing generates a vast number of read and write operations across a wide range of object sizes, from kilobytes (metadata headers) to multi-gigabyte tensors (optimizer states). This diversity leads to poor bandwidth utilization under traditional POSIX I/O. Emerging I/O libraries, such as \texttt{liburing}~\cite{axboe2022liburing}, provide a promising avenue to alleviate such inefficiencies by enabling batched, non-blocking data movement, combined with efficient shared buffers between user and kernel space to reduce per-call system overheads. When combined with file system flags like O\_DIRECT, or registered buffers, it can further enable zero-copy data movement between user-space and storage.
However, despite their potential, the efficiency kernel-assisted I/O libraries remains largely unexplored in the context of production-grade checkpoint/restore (C/R) engines. Crucial aspects such as I/O concurrency levels, buffered versus direct I/O, aggregation strategies need to be characterized in depth for successful adoption.

\paragraph{\textbf{Key Contributions.}}
Motivated by these challenges, this work conducts an extensive investigation into the I/O behavior of LLM checkpoint/restore. In particular, we focus on \texttt{liburing},a well established kernel-assisted I/O library integrated with the Linux kernel. Our findings aim to guide both the LLM and big data communities toward more performant, scalable, and data-aware C/R frameworks for next-generation AI workloads. We summarize our contributions as follows:

\begin{enumerate}[topsep=0pt,noitemsep,leftmargin=*]
    \item \textbf{A benchmarking framework for modern I/O engines: } We develop a microbenchmark designed to evaluate the performance of \texttt{liburing}, a kernel-accelerated I/O library, under both synthetic and realistic LLM workloads. The benchmark addresses limitations of existing tools by modeling \textit{file-per-shard} I/O-- the default behavior of modern checkpointing frameworks-- and capturing heterogeneous object sizes across ranks. It characterizes peak throughput and latency for emerging kernel-accelerated interfaces while systematically exploring how file system flags, aggregation strategies, and I/O concurrency affect checkpoint read and write performance at scale.

    \item \textbf{Characterizing aggregation for large LLM states: } Using this benchmark, we analyze I/O performance under varying aggregation granularities, bypass of caches, and data sizes. Our results show how grouping multiple tensors or model shards into larger I/O submissions affects throughput and metadata contention, exposing trade-offs between application-level concurrency and storage-level efficiency.

    \item \textbf{Evaluation of LLM checkpointing engines:} We empirically assess two state-of-the-art checkpointing runtimes-- PyTorch's TorchSnapshot~\cite{TorchSnapshot} and DataStates-LLM~\cite{Maurya_2024} through our microbenchmarks, highlighting how I/O patterns introduced by different phases of checkpoint and restore cycles, e.g., heterogeneity of data structures, buffering, (de)/serialization, and fragmented or uncoalesced I/O impact scalability on large parallel file systems.
\end{enumerate}

\section{Background and Related Works}
\begin{figure*}[!htb]
\centering
\minipage{0.20\textwidth}
    \centering
    \includegraphics[width=\linewidth]{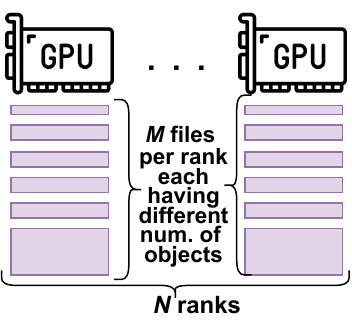}
    \Description{Infographic on how LLM checkpoints are broken down into multiple uneven files per GPU}
    \caption{Composition of a single LLM Checkpoint.}
    \label{fig:ckpt-struct}
\endminipage
\hfill
\minipage{0.42\textwidth}
    \centering
    \includegraphics[width=\linewidth]{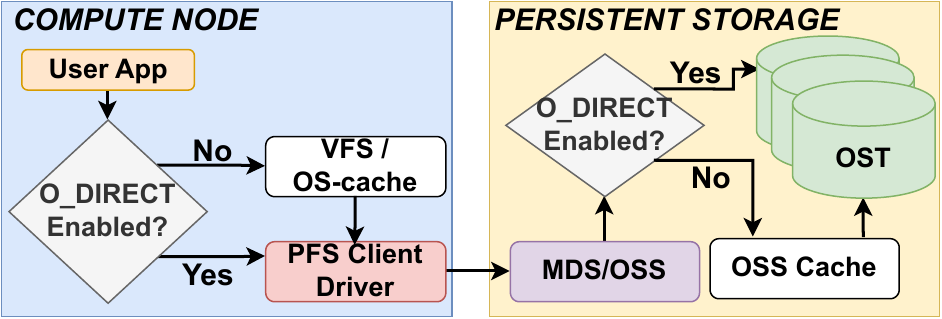}
    \vspace{10pt}
    \Description{flow diagram of data movement from user space to storage}
    \caption{Flow of Operations Based on Bypass Cache (O\_DIRECT) flags on the Parallel File System.}
    \label{fig:pfs-flow}
\endminipage
\hfill
\minipage{0.3\textwidth}
    \centering
    \includegraphics[width=\linewidth]{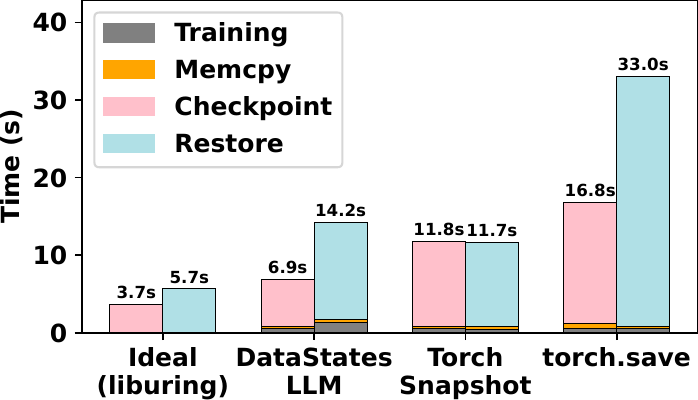}
   \vspace{-5pt}
   \Description{Bar chart comparing checkpoint and restore time across configurations, showing restore costs dominate total overhead}
    \caption{Checkpoint and Restore Overheads when Training a 3B model.}
     \label{fig:C-with3Bmodel}
\endminipage
\end{figure*}

\paragraph{\textbf{Composition of LLM Checkpoints.}}
\label{sec:checkpointstructure}
Modern LLMs generate vast amounts of data during training; for example, a 3B parameter model distributed across 4 GPUs produces tens of gigabytes of data per rank per checkpoint. While these data volumes are comparable to traditional HPC applications, the I/O characteristics differ substantially. In traditional settings, applications either checkpoint inherently large contiguous data structures or rely on C/R frameworks such as VELOC~\cite{veloc}, FTI~\cite{bautista2011fti}, and SCR~\cite{mohror2014scr} to pre-coalesce smaller structures into contiguous per-rank files on local storage before flushing to the PFS in fixed-size, aligned chunks and are largely not adopted for GPU-accelerated workloads. GPU-aware variants (e.g., VELOC-GPU~\cite{GPUPrefetch-HPDC23, VELOCGPU-HiPC22}, cudaCR~\cite{cudaCR}, CRIUgpu~\cite{CRIUgpu}, GDSCR~\cite{GDSCR}) extend this model to minimize device-to-host transfers via direct memory access, but still target large contiguous objects rather than the fragmented data typical of LLM checkpoints.

\begin{figure}[ht]
\begin{subfigure}{0.25\linewidth}
    \centering
    \includegraphics[width=\linewidth]{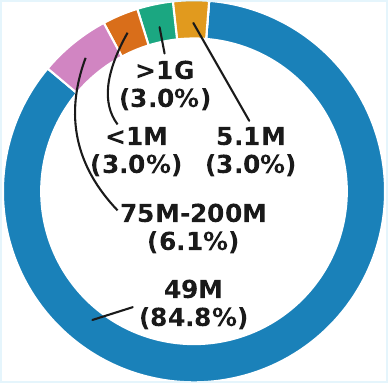}
    \Description{pie chart of file sizes per checkpoint iteration for standard BLOOM 3B model}
    \caption{3B model}
    \label{fig:3b-sizes}
\end{subfigure}
\hspace{5pt}
\begin{subfigure}{0.25\linewidth}
    \centering
    \includegraphics[width=\linewidth]{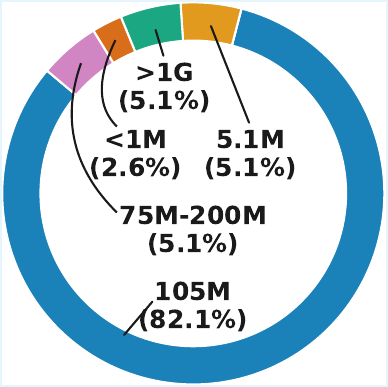}
    \Description{pie chart of file sizes per checkpoint iteration for standard Llama2 7B model}
    \caption{7B model}
    \label{fig:7b-sizes}
\end{subfigure}
\hspace{5pt}
\begin{subfigure}{0.25\linewidth}
    \centering
    \includegraphics[width=\linewidth]{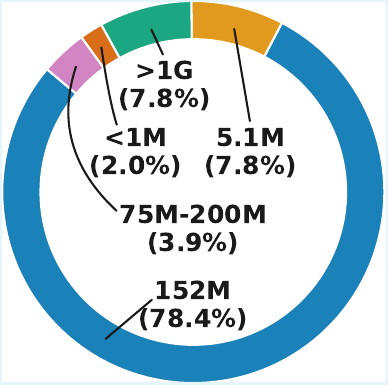}
    \caption{13 model}
    \label{fig:13b-sizes}
\end{subfigure}
\Description{pie chart of file sizes per checkpoint iteration for standard Llama2 13B model}
\caption{Checkpoint file size distribution of different models.}
\label{fig:ckpt-sizes-diff-models}
\end{figure}

Large models are typically sharded and parallelized across multiple GPUs using data-, tensor-, pipeline-parallelism, and ZeRO-redundancy-- collectively referred to as ``4D parallelism''~\cite{lian2025universalcheckpointingflexibleefficient}. While such strategies enables efficient training of LLMs~\cite{bloom}, it leads to a fragmented application state, wherein each GPU holds either replicated (data-parallel), or a unique shard (tensor, pipeline, ZeRO-redundancy) of the model or optimizer states~\cite{Maurya_2024}. To exploit parallel I/O, frameworks such as DeepSpeed~\cite{deepspeed4scienceinitiativeenablinglargescale} adopt a \emph{file-per-shard} checkpointing approach, resulting in an $N*M$ file layout that produces hundreds of files per rank. Furthermore, each file ranges from a few MB to multiple GB, as illustrated in Figure~\ref{fig:ckpt-struct}, and quantitatively detailed in Figure~\ref{fig:ckpt-sizes-diff-models}. While this ``application-centric'' layout simplifies runtime design, it dramatically increases file system metadata load. Even traditional HPC's file-per-process can generate significant bottlenecks and management challenges~\cite{VELOC-FGCS24, lustre}; LLMs exacerbate this issue by massively multiplying file counts, stressing both compute and storage resources.

Many LLM-specific C/R engines, such as DataStates-LLM~\cite{Maurya_2024} adopt this same checkpointing pattern, as it is simple to integrate with existing training frameworks. Other solutions, such as TorchSnapshot~\cite{lightning_async_checkpoint_io}, exacerbate this problem even more: large objects and model states are subdivided into fixed size amounts (512~MB by default), and each fixed-size chunk is flushed to a separate file inside a deeply nested subdirectory, stressing all levels of the PFS (MDS, OSS, and OSTs). Furthermore, these frameworks immediately issue write operations for each data structure within the checkpoint object, regardless of size, leading to higher MDS contention and suboptimal usage of available PFS bandwidth, which is better designed for large, contiguous transfers. These challenges highlight a key tension in LLM checkpoint design: while the application-centric layouts simplify integration with training frameworks, they diverge sharply from the filesystem-friendly patterns that underpin traditional HPC checkpointing, motivating the need for aggregation and smarter I/O orchestration to fully leverage PFS capabilities.

\paragraph{\textbf{I/O Path on the Parallel File Systems}}
\label{sec:storageflow}
Modern HPC systems connect compute nodes to a shared parallel file system (PFS) through high-bandwidth interconnects (e.g., InfiniBand). A PFS comprises a \emph{metadata server} (MDS) that manages filesystem metadata stored on \emph{metadata targets} (MDTs), and multiple \emph{object storage servers} (OSSs) that handle file data stored on \emph{object storage targets} (OSTs). Compute nodes interact with the MDS for metadata operations and with the OSSs for data I/O, as illustrated in Figure~\ref{fig:pfs-flow}.

At runtime, OS-level caches on compute and storage nodes act as transient burst buffers between the application and the PFS. The \texttt{O\_DIRECT} flag-- a POSIX file system flag specified during file open-- controls whether I/O bypasses these caches, enabling direct transfers between user space and storage, eliminating double-buffering at the cost of exposing raw storage latency. Conversely, cached I/O can mask latency and coalesce small writes, but may inflate memory use and obscure performance analysis.

These tradeoffs are central to understanding checkpoint I/O. Fine-grained, writes from hundreds of ranks trigger thousands of metadata and small I/O operations, stressing the PFS. Monitoring tools like Darshan~\cite{darshan} reveal these bottlenecks by exposing file access patterns and metadata load on production systems~\cite{saeedizade2023ioburstpredictionhpc}. However, Darshan’s analysis is limited for asynchronous I/O, as it records only request submission times rather than actual I/O completion, and currently lacks integration with interfaces such as \texttt{liburing}. This gap underscores the need for new performance models and measurement tools that reflect LLM checkpoint/restore workloads on shared storage systems.

\paragraph{\textbf{Kernel Accelerated I/O Libraries}}
The emergence of kernel-bypass and low-overhead I/O frameworks, such as \texttt{io\_uring}, \texttt{libaio}, and \texttt{SPDK}~\cite{ren2023performance, didona2022understanding}, introduces new opportunities to optimize data paths in modern storage systems. Libraries like \texttt{liburing} expose submission and completion queues directly in user space, enabling batched, asynchronous I/O and reducing system call and context switch overhead. While this does not inherently bypass the file system or enable “direct-to-storage” transfers (which still depend on \texttt{O\_DIRECT}), it eliminates much of the user–kernel transition cost that traditionally limits small, frequent I/O operations~\cite{didona2022understanding}.

In the context of LLM checkpoints, such kernel-accelerated interfaces allow applications to pipeline data movement between user-space buffers and storage asynchronously, breaking the bottleneck of sequential I/O and improving host CPU utilization. When combined with aggregation and multithreaded flushing, \texttt{liburing} can approach or even exceed the throughput of well-tuned POSIX I/O at scale, particularly under high concurrency~\cite{axboe2022liburing}.

However, leveraging these advantages requires careful alignment between the application’s internal I/O granularity and the PFS’s stripe and metadata behavior. Without such coordination, asynchronous I/O may increase contention or induce small, misaligned writes. Consequently, an ideal checkpointing system must not only overlap computation and I/O but also co-design its buffering, aggregation, and kernel-level I/O strategies with the underlying storage system to fully exploit available bandwidth.

\paragraph{\textbf{Dissecting The Flow of Events During Checkpoint and Restore Operations}}
As depicted in Figure~\ref{fig:ckpt-struct}, each checkpoint file represents a logical object composed of multiple nested data structures-- strings, runtime class objects, random state iterators, lists, tensors, etc. The major size of each file is dictated by tensors, which reside both on the CPU and the GPU. Unlike other data structures, which need to be serialized and converted to a byte stream to be stored on disk, tensors and contiguous containers (e.g., numpy arrays) are inherently contiguous byte streams, i.e., pre-serialized buffers, and can be saved directly. Therefore, due to large sizes, distributed memory locations (GPU and host memory), and zero serializations needed, tensors and the remaining data structures are treated separately during checkpoint/restore, as detailed next.

For each logical checkpoint object, the default checkpointing approach-- \texttt{torch.save}, adopted by training runtimes~\cite{rasleyDeepSpeedSystemOptimizations2020}, synchronously and sequentially allocate host memory for all GPU resident data structures, transfer them from GPU to the host memory, serialize the entire logical object, and finally flush to disk, leading to significant I/O overheads, often dominating the training time~\cite{Maurya_2024}. To accelerate checkpointing, state-of-the-art LLM engines such as TorchSnapshot~\cite{TorchSnapshot} and DataStates-LLM~\cite{Maurya_2024} decompose the checkpoint operation on each of the $N*M$ files into several synchronous and asynchronous stages:

\begin{enumerate}[topsep=0pt,itemsep=0pt,leftmargin=12pt]
    \item \textbf{Tensor extraction and ``lean object'' serialization:} Tensors, both resident on GPU or host memory, are detached from the logical checkpoint object, and the remaining data, referred to as ``lean checkpoint object'' is serialized (typically using \texttt{pickle} module). Given the Global Interpreter Lock (GIL) limitations of Python<3.14, this stage was synchronous, i.e., blocked training.
    \item \textbf{GPU to host transfers:} GPU resident datastructures, typically tensors, are flushed on (pre-allocated) host buffers, either synchronously (TorchSnapshot~\cite{TorchSnapshot}), or asynchronously (DataStates-LLM~\cite{Maurya_2024}), i.e., overlapped transfers with immutable forward and backward phases of training.
    \item \textbf{Host to file transfers:} All tensors, serialized datastructures, and metadata of the logical checkpoint object are asynchronously flushed to slower storage tiers while training proceeds.
    \item \textbf{Header/metadata generation:} Metadata headers map tensors to offsets in files for reconstruction during the restore.
\end{enumerate}

During restore, all the $N$ processes operate in parallel and read back their $M$ files each from the latest checkpoint version directory. While the $N$ processes restore in parallel, all checkpoint engines restore the $M$ logical objects (or files) serially, i.e., the next file is read only when the previous logical checkpoint object has been fully restored on both GPU and host tiers. Unlike checkpointing, wherein some stages are asynchronous, during restore, all checkpointing engines adopt a synchronous and serial read approach, i.e., each data structure of each logical object on a given process is moved across one tier at a time, and only when restored on the destination memory, the next data structure is read from the file.

Conventional approaches, i.e., \texttt{torch.load}, opaquely allocate memory for the entire checkpoint object, read it from the file, perform deserialization of the entire object, and transfer data structures to the GPU if needed. However, due to the separation of pre-serialized tensors and the ``lean checkpoint object'' during checkpointing, state-of-the-art engines cannot adopt such opaque policies, and restore using the following stages.

\begin{enumerate}[topsep=0pt,itemsep=0pt,leftmargin=12pt]
    \item \textbf{Reading the metadata} triggers a few bytes of read operation.
    \item \textbf{Restoring the ``lean checkpoint object''}, based on the offset dictated by the metadata header, the ``lean checkpoint object'', typically of a few kilobytes, is read back and de-serialized.
    \item \textbf{Restoring tensors} by iterating over each tensor entry in metadata, host memory is allocated, where the tensor is copied from the file, triggering reads amounting to tens of MB to tens of GB.
    \item \textbf{Host to GPU transfers} are triggered once the entire ``checkpoint object'' is reconstructed on the host memory, after which the GPU resident data structures are copied from host to GPU synchronously.
\end{enumerate}

This design, while modular and simple, yields inefficient I/O behavior. Restoring thousands of small objects and fragmented files lead to high metadata traffic and low effective throughput, as each tensor, metadata entry, or manifest component often triggers an independent read or write call. Furthermore, both frameworks perform separate reads for metadata and data objects. TorchSnapshot first reads a single manifest file describing all tensors and metadata, then iteratively restores each object one-by-one—triggering one read call per object. In contrast, DataStates-LLM issues a separate read for every entry referenced in the metadata header, including one for the metadata itself, one for the serialized (“lean”) object, and one for each tensor or chunk—typically tripling the number of read operations. Furthermore, both frameworks allocate host memory for each read on the fly rather than preallocating buffers, introducing additional memory management overhead and preventing efficient pipelining of I/O and data placement.

Together, these design choices: per-object granularity, lack of I/O coalescing, and dynamic memory allocation create substantial metadata contention and underutilize available bandwidth. They highlight a broader tension between modular, application-centric checkpointing and filesystem-efficient I/O, motivating the need for aggregation and coordinated data movement strategies that consolidate small reads/writes into fewer, larger transfers.

\paragraph{\textbf{Motivation}}
To understand the I/O overheads during LLM checkpoint and restore, we consider the case of 3B model, running on 4$\times$A100-40GB GPUs using tensor-parallelism, meaning each model layer is partitioned across the GPUs, and each GPU holds a distinct shared of the model's weights and activations. Cumulatively across the 4 ranks, the 3B model produces 132 files, aggregating to 42~GB for each checkpoint. We study the I/O overheads of fully persisting a checkpoint to the PFS, and then restoring it using three engines: default deepspeed (\texttt{torch.save}), Pytorch's TorchSnapshot~\cite{TorchSnapshot}, and DataStates-LLM~\cite{Maurya_2024}. In addition to the LLM engines, we also compare against an ``ideal approach'', simulated through a microbenchmark, which flushes from 4 ranks, the same volume of a single large host-resident, contiguous data buffer using the \texttt{liburing} library. We profile the time taken by the major phases of an iteration, i.e., forward, backward, device transfers, and checkpoint/restore times (data loading and update phase are negligible).

\paragraph{\textbf{Gaps in Checkpointing:}}
The pink bars in Figure~\ref{fig:C-with3Bmodel} show the breakdown of a single iteration when checkpointing using different engines. When compared against the ``ideal approach'', we observe that DataStates-LLM, TorchSnapshot, and default \texttt{torch.save} checkpoint engines lead to 1.8$\times$, 3.2$\times$, and 4.5$\times$ slower iterations, respectively. This hints at significant I/O gaps in LLM checkpointing engines, which has not been previously characterized and studied.

\paragraph{\textbf{Gaps in Restore:}}
Next, we study the breakdown of a single iteration when restarting, depicted using the blue bars in Figure~\ref{fig:C-with3Bmodel}. The TorchSnapshot engine performs the fastest restoration of the checkpoint, outperforming DataStates-LLM and the default \texttt{torch.save} approach by 1.22$\times$ and 2.8$\times$, respectively. Similar to the case of checkpointing, we observe that all approaches lag behind the ``ideal approach'' by at least 51\%, showing potential for accelerating restore phases. However, despite the common notion of reads being faster than writes, on our platform~\ref{sec:expt:testbed}, we observe a reversed trend wherein checkpoint flushes are faster than restore reads.

These findings highlight a growing disconnect between the checkpointing abstractions used by LLM frameworks and the ideal access patterns expected by high-performance file systems. \emph{Bridging this gap requires rethinking how model states are serialized, grouped, and issued to the storage stack to fully exploit available I/O bandwidth}.

\section{Study of I/O Patterns and PFS Interactions}

\subsection{Experimental Testbed}
\label{sec:expt:testbed}
We conduct our experiments on the ALCF Polaris System, comprising 560 nodes each equipped with 1 AMD EPYC Milan 753P CPU, 4$\times$A100-40GB GPUs, and 512 GB of DDR4 with a bandwidth of 204.8 GB/s. The system is connected to a Lustre-based PFS with 100 PB of storage, containing 40 metadata servers~(OSSes) and 160 object storage targets~(OSTs), with an aggregate bandwidth of 650 GB/s. We change the default stripe settings to allow all files to be striped across all available OSTs and set the stripe size to 64 MB, which better ensures that concurrent requests are spread out among more OSTs and each I/O operation touches 1 OST. The microbenchmark uses liburing version 2.12 and is compiled with gcc 13.3.1.

We present a benchmarking framework that captures both synthetic and LLM-realistic workloads to evaluate I/O performance under varying aggregation strategies and modern I/O interfaces such as \texttt{liburing} and \texttt{libaio}. By systematically varying tensor sizes and aggregation layout, our framework isolates performance bottlenecks in the checkpoint/restore pipeline and provides a detailed performance model of memory- and storage-level interactions during C/R I/O. This approach enables precise characterization of serialization, batching, memory limitations, and metadata overheads that affect overall throughput.

\subsection{Methodology}
This work aims to characterize peak performance of the Linux \texttt{liburing} interface under C/R workloads representative of LLM training and compare how close state-of-the-art C/R engines approach this peak. We design a suite of experiments that establish performance models for \texttt{liburing} under a range of I/O conditions: ring contention levels, checkpoint sizes, scaling the number of processes performing checkpointing, aggregation granularity, and caching modes. These experiments isolate how kernel-level I/O acceleration behaves under LLM's multi-file and concurent checkpointing requirements, helping identify design principles for future checkpointing engines that better align with PFS capabilities.

\subsubsection{\textbf{I/O Aggregation and Data Locality}}
Aggregation plays a critical role in determining both I/O throughput and PFS scalability~\cite{VELOC-FGCS24}. Fine-grained, file-per-object strategies used in current LLM frameworks create excessive metadata contention and make suboptimal use PFS bandwidth, whereas coarser aggregation improves data locality and aids I/O coalescing, but risks coordination overhead and OST contention.

We explore this trade-off through three representative strategies:
\begin{enumerate}[topsep=0pt,itemsep=0pt,leftmargin=12pt]
    \item \textbf{File-per-Tensor:} Each tensor or shard is written independently, reflecting the uncoalesced I/O patterns of frameworks such as DeepSpeed and TorchSnapshot and exposes the metadata and queue contention costs at scale
    \item \textbf{File-per-Process:} Each rank writes all its data to a single file, reducing file handle count and improving sequential locality. This strategy demonstrates the benefits of moderate aggregation for typical PFS configurations.
    \item \textbf{File-per-Tensor:} All ranks write to disjoint, contiguous offsets in a shared file. This represents the best-case scenario for data coalescing and enables us to model contention effects of both the PFS and submission ring and completion queues of \texttt{liburing}.
\end{enumerate}

By studying these strategies, we characterize how \texttt{liburing} handles aggregation compared to traditional POSIX I/O, which is known to degrade rapidly under high file counts but has not been extensively evaluated in kernel-accelerated contexts.

\subsubsection{\textbf{Buffered vs. Direct I/O Paths}}
We examine how the data path impacts performance by enabling and disabling the \texttt{O\_DIRECT} flag, which bypasses the Linux page cache. Buffered I/O routes data through the kernel’s caching layer, potentially improving small sequential writes but introducing additional copies and unpredictable flushing. In contrast, \texttt{O\_DIRECT} transfers data directly between user buffers and storage, eliminating caching overhead but requiring strict alignment and padding to the filesystem block size.
This experiment quantifies how caching affects throughput asymmetrically for writes and reads, especially under mixed object sizes. POSIX serves as the baseline here to contrast user-space buffering behavior against \texttt{liburing}’s kernel-bypass semantics.

\subsubsection{\textbf{Benchmark Design}}
We design two complementary benchmarks to provide both controlled performance models and LLM-representative evaluations.

\begin{enumerate}[topsep=0pt,itemsep=0pt,leftmargin=12pt]
 \item {\textbf{Synthetic Benchmark}}:
This benchmark isolates intrinsic I/O characteristics without higher-level framework overheads (fragmentation, uneven object sizes, non-aligned offsets). Each rank issues contiguous writes from host-resident buffers of 128~MB–8~GB, approximating the total checkpoint volumes of mid-scale models while maintaining controlled granularity.

\item {\textbf{Representative LLM Benchmark}}:
\label{sec:LLM-bench-design}
The second benchmark models realistic checkpoint workloads derived from BLOOM-3B, LLaMA-7B, and LLaMA-13B training configurations outlined in Figure~\ref{fig:ckpt-sizes-diff-models}. Checkpoint structures reproduce the actual file layouts, tensor distributions, and process counts used during training (4, 8, and 16 ranks respectively). Each rank manages multiple buffers representing model and optimizer states and issues I/O batches up to the configured queue depth before awaiting completions, capturing non-uniform and staggered submission behavior.
We use this benchmark to evaluate how \texttt{liburing} behaves under real-world fragmentation, varying aggregation strategies, and different data paths.
\end{enumerate}

\subsection{I/O Coalescing with Synthetic Benchmark}
We first evaluate \texttt{liburing} in isolation using the synthetic benchmark. In both experiments there are no more than 4 processes per node (matching Polaris' GPU setup). Each rank divides its large buffer into 64 MB regions, similar to DataStates-LLM~\cite{Maurya_2024}, and submits all chunks at once, enabling us to evaluate \texttt{liburing}’s concurrent I/O. Varying the total buffer size allows us to explore both how liburing handles data sizes representative of checkpoint objects and the maximum concurrency the queue or PFS can sustain. Inspired by DataStates-LLM, which coalesces objects into host buffers but submits I/O as soon as each object is available, our benchmark instead accumulates data up to the buffer size before issuing a batched flush, balancing minimized application stalls with high sustained PFS throughput.

\begin{figure*}[!htb]
\minipage{0.32\textwidth}
    \centering
    \includegraphics[width=\linewidth]{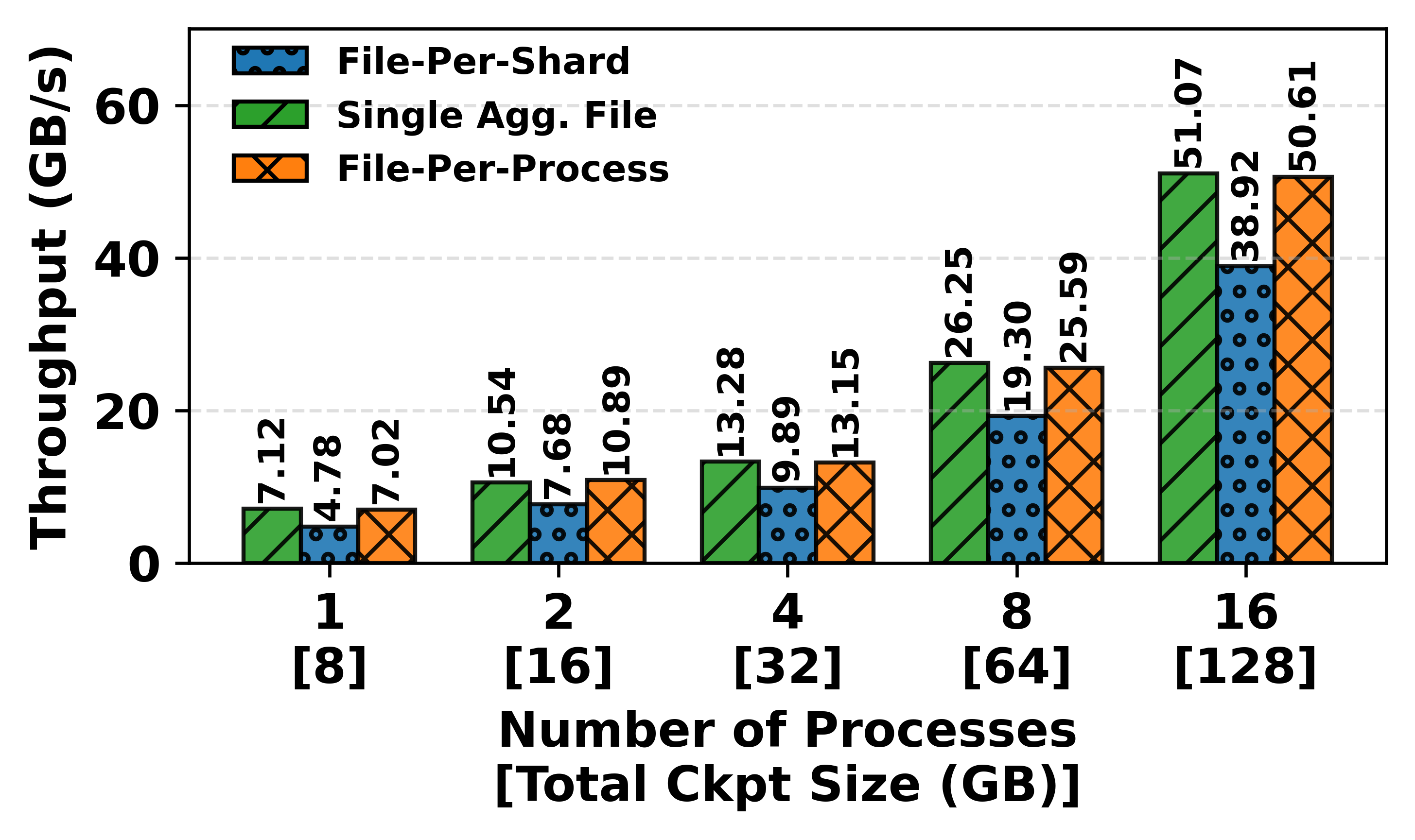}
    \Description{bar chart comparing write performance for different levels of aggregation on the synthetic benchmark, using 1-16 processes and showing more aggregated solutions obtaining better throughput}
    \caption{Write throughput comparing the 3 aggregation strategies on the synthetic benchmark (1 - 16 processes with up to 4 processes per node where each process checkpoints 8 GB of data; higher is better)}
    \label{fig:synthetic_aggregation_multinode_write}
\endminipage
\hfill
\minipage{0.32\textwidth}
    \centering
    \includegraphics[width=\linewidth]{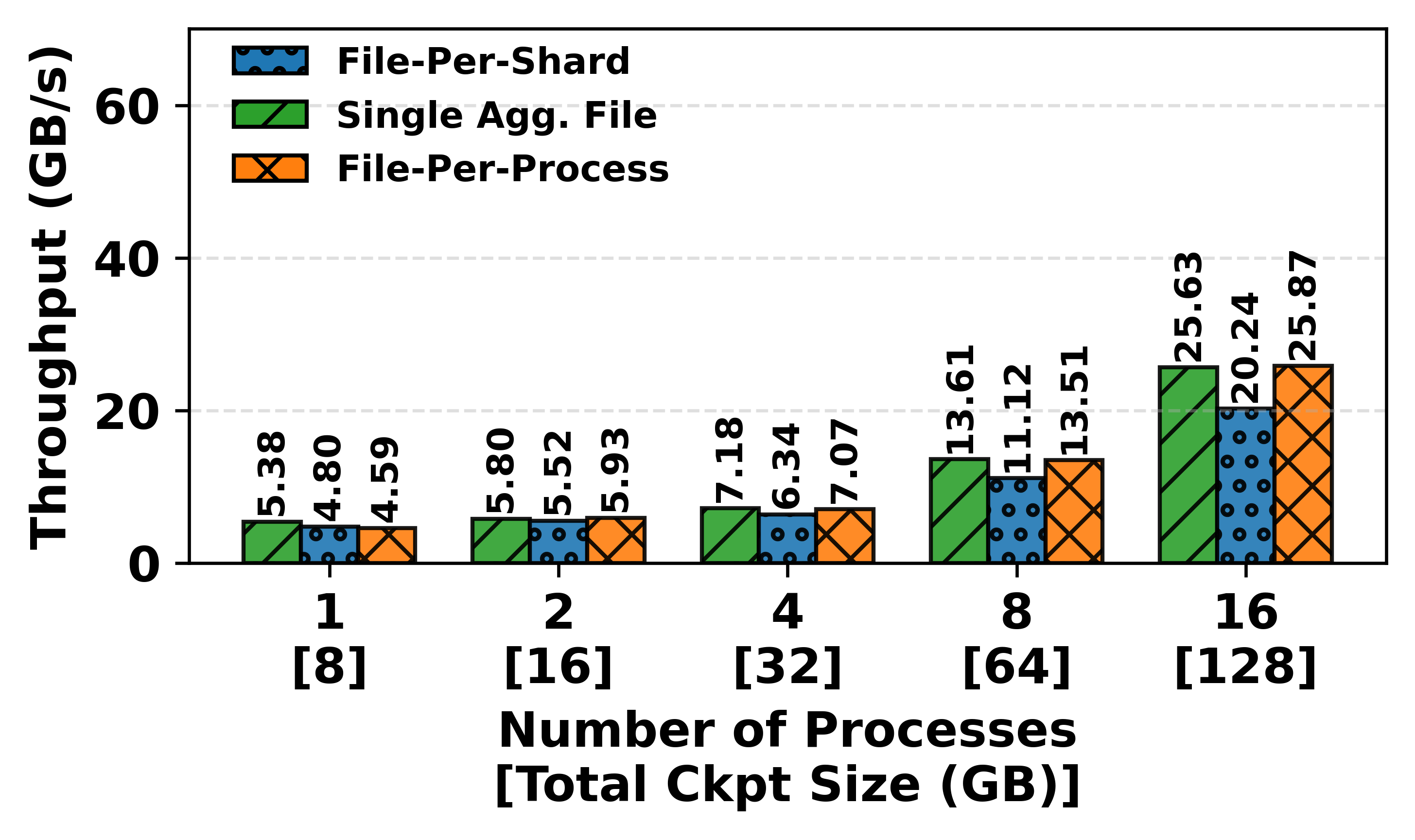}
    \Description{bar chart comparing read performance for different levels of aggregation on the synthetic benchmark, using 1-16 processes and showing more aggregated solutions obtaining better throughput}
    \caption{Read throughput comparing the 3 aggregation strategies on the synthetic benchmark (1 - 16 processes with up to 4 processes per node where each process restores 8 GB of data; higher is better)}
    \label{fig:synthetic_aggregation_multinode_read}
\endminipage
\hfill
\minipage{0.32\textwidth}
    \centering
    \includegraphics[width=\linewidth]{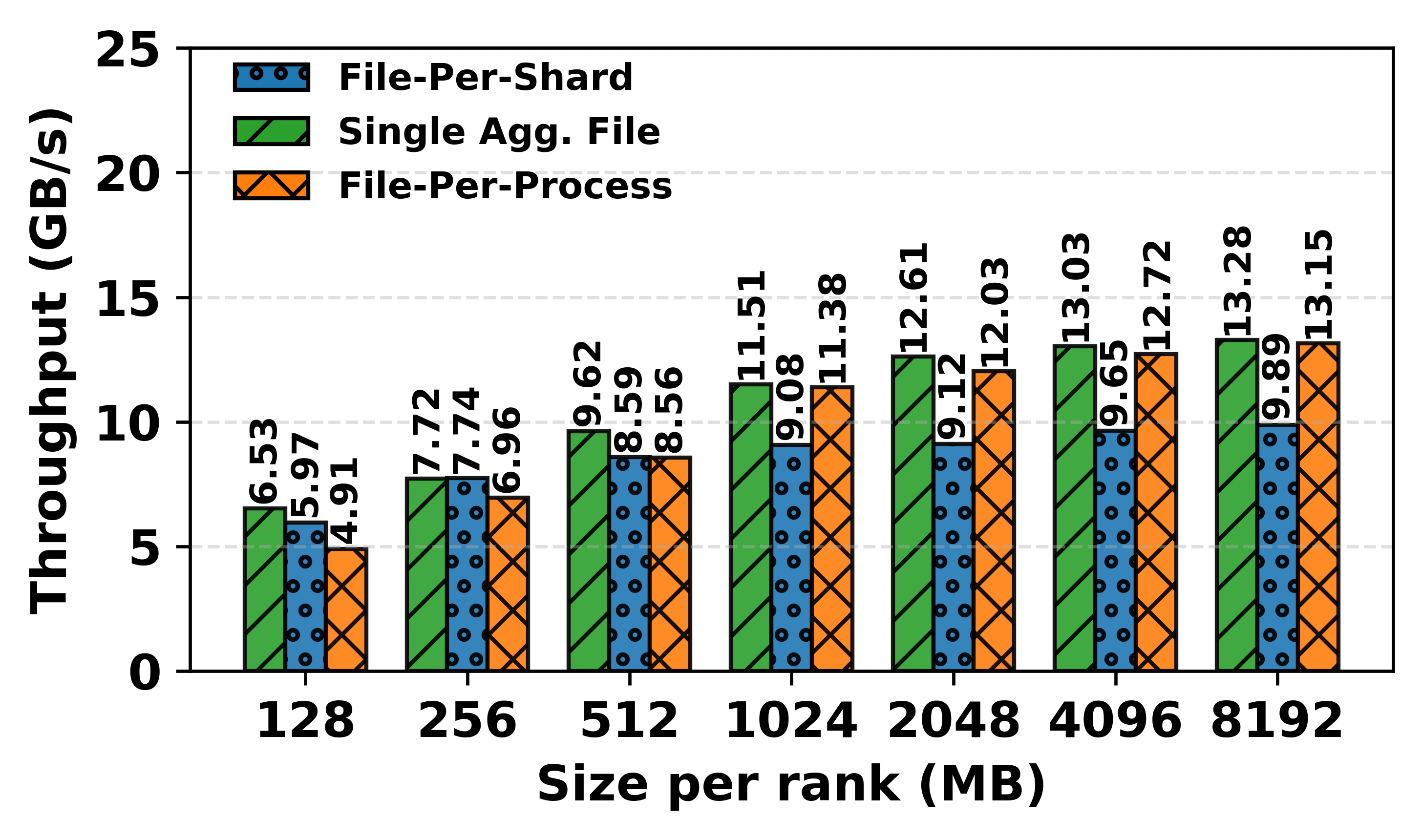}
     \Description{bar chart comparing write performance for different levels of aggregation on the synthetic benchmark, varying tensor sizes by powers of 2 and showing more aggregated solutions obtaining better throughput}
    \caption{Write throughput comparing the 3 aggregation strategies for the synthetic benchmark varying the contiguous tensor size between 128~MB-8~GB (1 compute node using 4 processes; higher is better)}
    \label{fig:synthetic-agg-singlenode-writes}
\endminipage
\hfill
\minipage{0.32\textwidth}
    \centering
    \includegraphics[width=\linewidth]{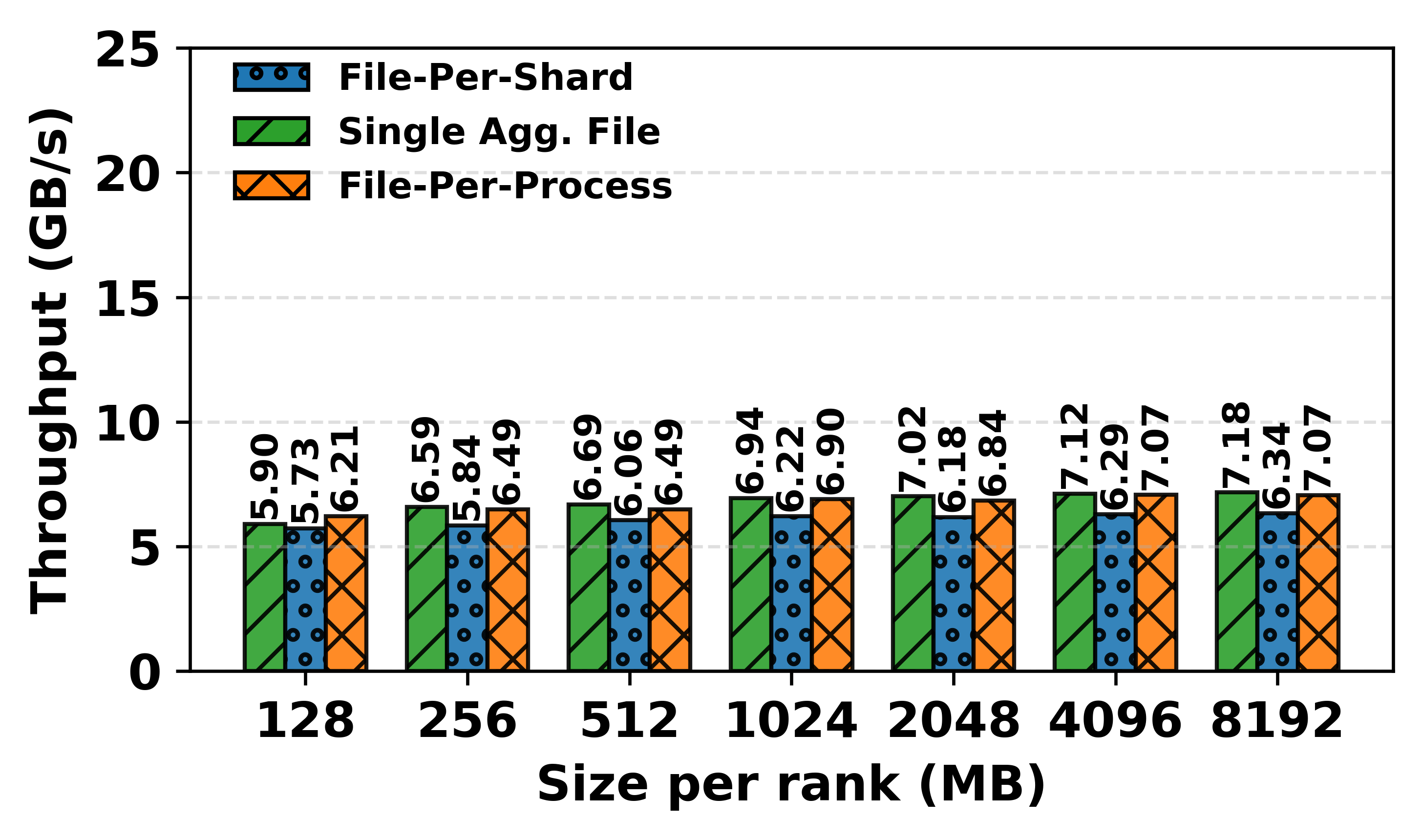}
    \Description{bar chart comparing read performance for different levels of aggregation on the synthetic benchmark, varying tensor sizes by powers of 2 and showing more aggregated solutions obtaining marginally better throughput but not as much as the write throughput}
    \caption{Read throughput comparing the 3 aggregation strategies for the synthetic benchmark varying the contiguous tensor size between 128~MB-8~GB (1 compute node, using 4 processes; higher is better)}
    \label{fig:synthetic-agg-singlenode-reads}
\endminipage
\hfill
\minipage{0.32\textwidth}
    \centering
    \includegraphics[width=\linewidth]{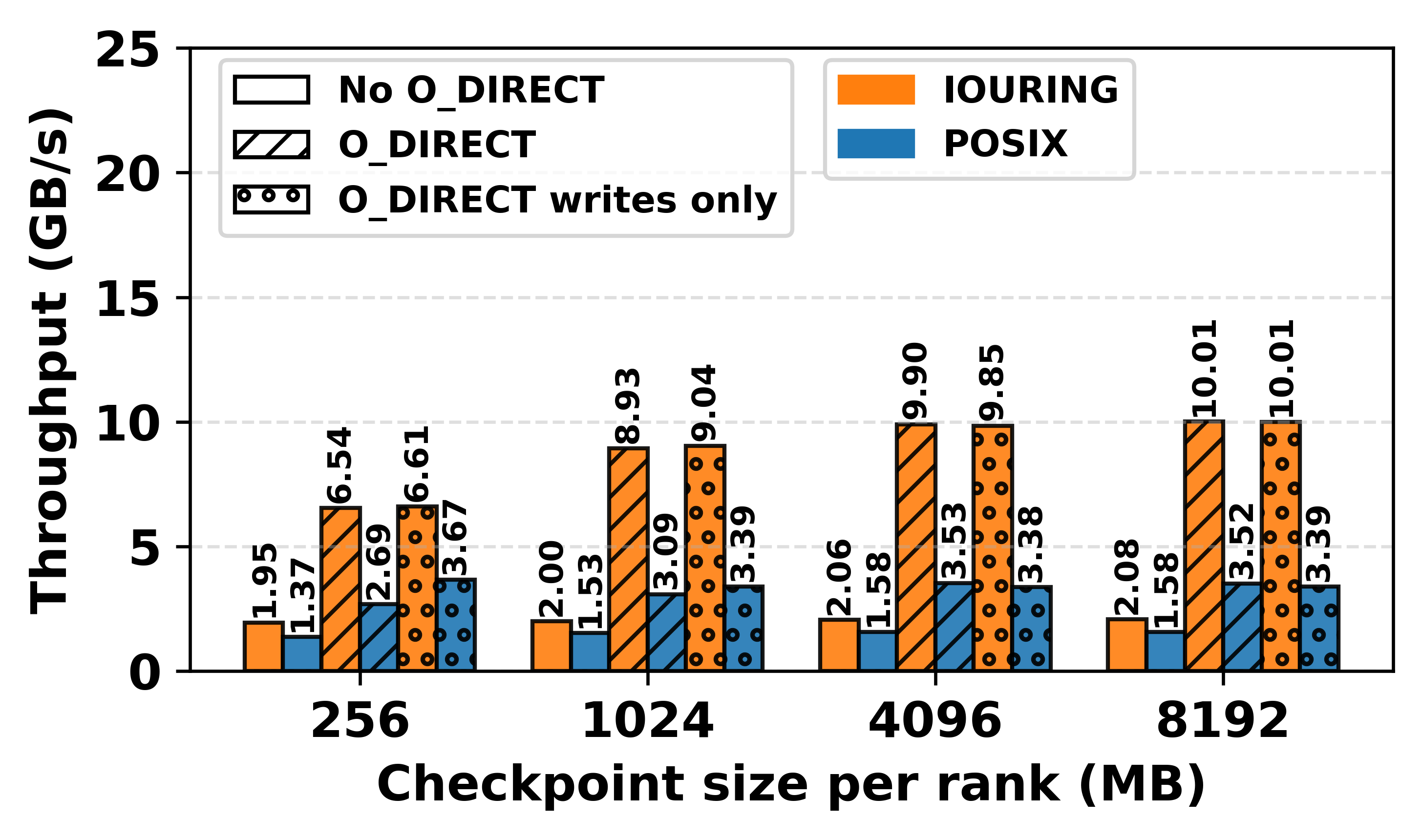}
    \Description{bar chart comparing POSIX and libruing write performance across varying tensor sizes and under different O\_DIRECT configurations where O\_DIRECT enabled improves write throughput for both strategies, and liburing approximately 2 times better than POSIX}
    \caption{Write throughput varying data region sizes in multiples of 4 from 256M - 8G for both POSIX and liburing under different O\_DIRECT flag configurations. (1 compute node using 4 processes; higher is better).}
    \label{fig:odirect-writes}
\endminipage
\hfill
\minipage{0.32\textwidth}
    \centering
    \includegraphics[width=\linewidth]{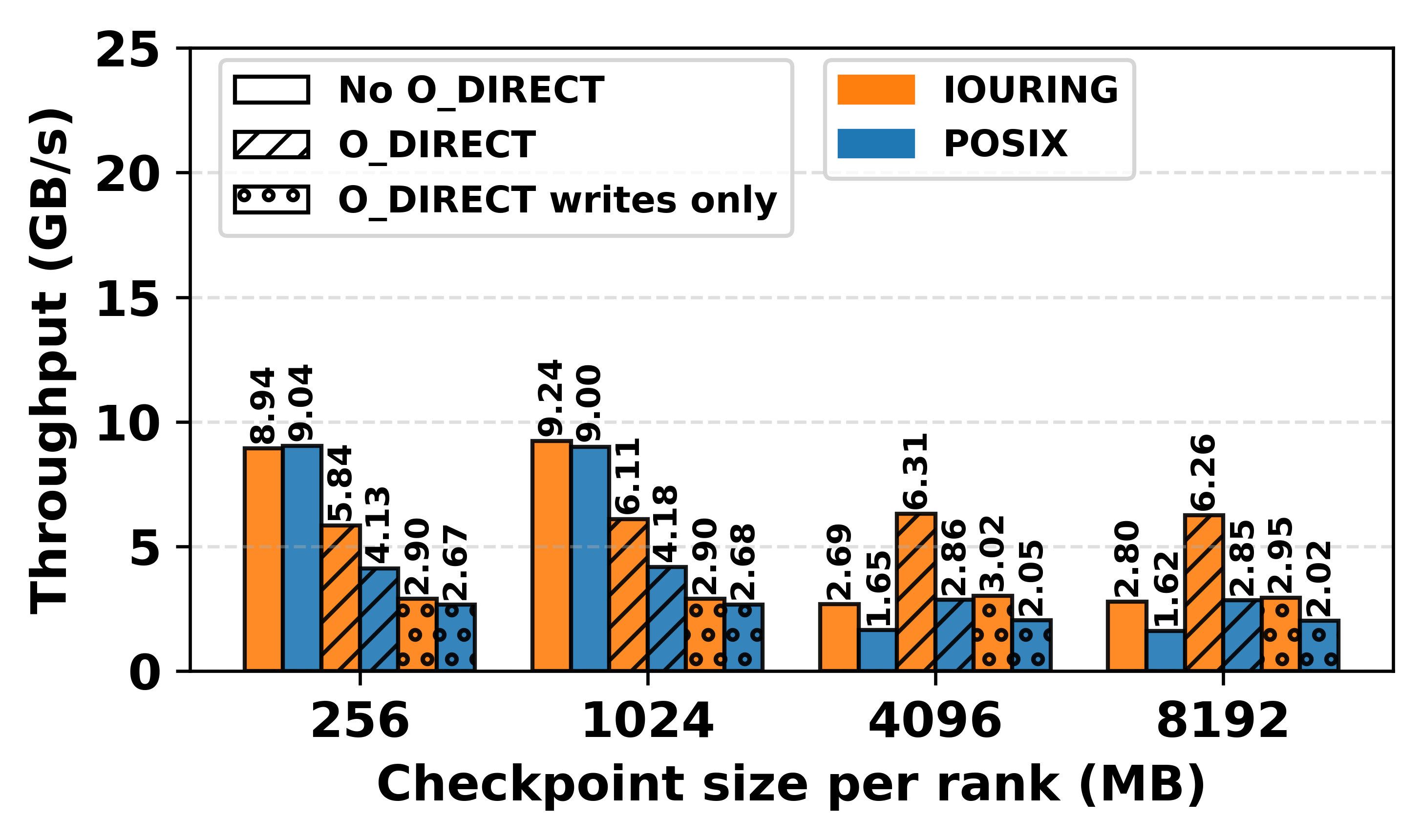}
    \Description{bar chart comparing POSIX and libruing read performance under across varying tensor sizes and different O\_DIRECT configurations where O\_DIRECT disabled improves read throughput for both strategies, with the notable exceptions at large tensor sizes (4 and 8 GB)}
    \caption{Read throughput varying data region sizes in multiples of 4 from 256M - 8G for both POSIX and liburing under different O\_DIRECT flag configurations. (1 compute node using 4 processes; higher is better).}
    \label{fig:odirect-read}
\endminipage
\end{figure*}

Figures~\ref{fig:synthetic_aggregation_multinode_write} and~\ref{fig:synthetic_aggregation_multinode_read} illustrate the I/O throughput during writes and reads on a small scalability experiment, varying the number of processes from 1-16, with 4 processes per node. Each process checkpoints and restores 8 GB of data (the largest file size in the BLOOM-3B model).
Both aggregation strategies (even a naive write-at-offset), outperform the file-per-shard strategy by up to $\approx$34\%. File-per-process and single aggregated file strategies show similar throughput, suggesting that liburing suffers contention as the number of files increases: multiple files require repeated kernel lookups, permission checks, block I/O setup, and lock management, whereas a single file allows cached reuse and reduces overhead. Read throughput on the other hand remains relatively stagnate across 1-4 processes, suggesting the node-level outgoing bandwidth is capped around 7 GB/s.

Figures~\ref{fig:synthetic-agg-singlenode-writes} and~\ref{fig:synthetic-agg-singlenode-reads} zoom in on single-node performance (4 processes) across data sizes from 128~MB to 8~GB. Write throughput scales with per-rank data size up to $\approx$2 GB before plateauing as node resources saturate, while read throughput stays roughly constant and $\approx2\times$ lower than writes. Aggregation consistently outperforms file-per-tensor layouts across all configurations, confirming its effectiveness in improving I/O efficiency.

\vspace{10pt}
\fcolorbox{black}{purple!8}{%
  \begin{minipage}{0.95\linewidth}
  {\bf Observation 1:} Aggregation is shown to be necessary for sustaining throughput in LLM checkpointing. In multi-node experiments (figures~\ref{fig:synthetic_aggregation_multinode_write} and~\ref{fig:synthetic_aggregation_multinode_read}) liburing exhibits increasing contention as the number of per-rank files grows, causing throughput to degrade even when aggregate I/O volume remains constant. Single-file aggregation mitigates this effect by reducing metadata presure and kernel-level coordination overheads that are amplified by the highly sharded checkpoint layouts of modern LLMs. On a single-node, write throughput scales only up to approximately 2 GB per rank before saturating, indicating a practical upper bound on effective batching. This saturation point defines an LLM-specific aggregation threshold that training frameworks such as DataStates-LLM must target to maximize sustained PFS utilization without incurring diminishing returns.
  
  Aggregation improves throughput across all configurations. Multi-node experiments show that liburing experiences contention when managing numerous files, favoring single-file aggregation. Single-node tests reveal that write throughput scales up to $\approx$2 GB per rank before saturating, suggesting an optimal batching threshold for frameworks like DataStates-LLM to maximize sustained PFS performance.
  \end{minipage}
}

\subsection{O\_DIRECT Influence on Writes and Reads}
Figures~\ref{fig:odirect-writes} and~\ref{fig:odirect-read} evaluate the impact of the \texttt{O\_DIRECT} flag on \texttt{liburing} and POSIX throughput using the synthetic benchmark. We use 4 processes on a single node and focus on the single aggregated-file configuration, varying data sizes from 256~MB to 8~GB to match representative checkpoint sizes. POSIX is included as a baseline to distinguish whether observed effects stem from system-level caching or \texttt{liburing} itself.

\texttt{O\_DIRECT} bypasses both kernel page caching and OSS-level buffering, forcing data to move directly between user-space buffers and storage. This change affects reads and writes in opposing ways due to differences in how the kernel manages cache residency and delayed flushing. For writes, avoiding double buffering eliminates extra memory copies and writeback synchronization, yielding up to $4.8\times$ and $2.2\times$ higher throughput for \texttt{liburing} and POSIX, respectively. The larger benefit observed with \texttt{liburing} reflects its lower system call overhead and the ability to batch I/O submissions without invoking kernel-side buffering mechanisms. In contrast, reads degrade when caching is disabled because the kernel cannot exploit page cache locality or readahead to prefetch contiguous regions. Buffered reads remain up to $2.3\times$ faster for smaller data sizes ($\leq$1~GB), where repeated accesses can be served directly from memory, but this advantage disappears once client and OSS caches saturate beyond $\approx$4~GB. At this point, \texttt{O\_DIRECT} achieves slightly higher and more stable throughput, as it avoids cache eviction and memory pressure across multiple processes issuing concurrent reads.

Therefore, a natural solution would seem to be to use the O\_DIRECT flag only for writes and disable it for reads when the data region is $\leq \approx1$ GB. However, as shown in figure~\ref{fig:odirect-read}, this yields no improvement, even further reducing throughput by up to $3\times$ for liburing and $3.3\times$ for POSIX. When buffering is reintroduced for large, cold reads, the kernel and OSS attempt to cache data opportunistically but cannot sustain it at scale, leading to the same disk-bound performance with added cache management overhead. For these reasons, we opt to keep the O\_DIRECT flag enabled for both writes and reads throughout the remainder of our experiments, as it provides the best balance between write and read performance.

\begin{figure*}[ht]
\minipage{0.32\textwidth}
    \centering
    \includegraphics[width=\linewidth]{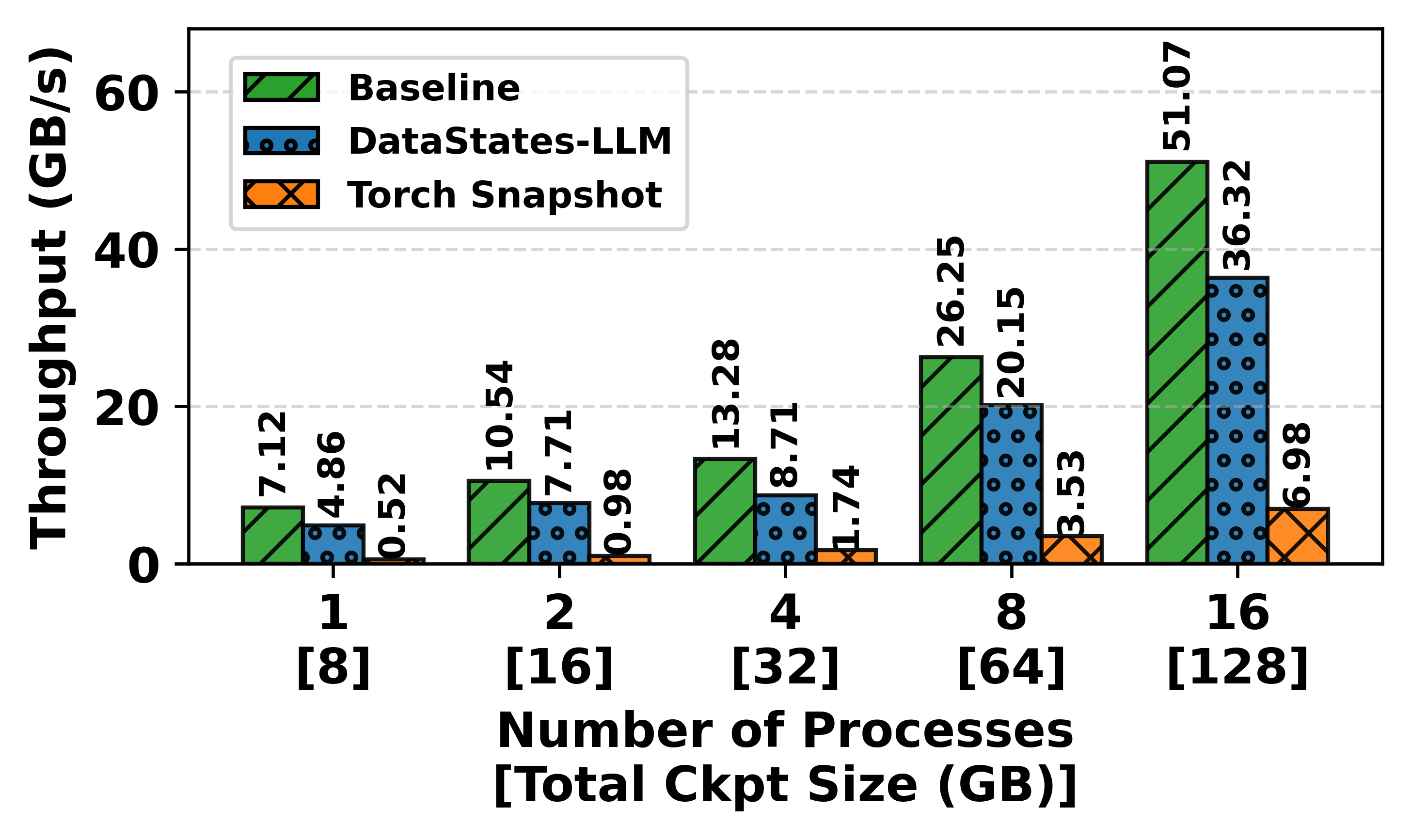}
    \Description{bar chart comparing write performance using different checkpointing engines (DataStates-LLM and TorchSnapshot) to the liburing baseline on the synthetic benchmark. The liburing baseline outperforms both checkpointing engines}
    \caption{Checkpoint throughput of the synthetic benchmark using the single aggregated file configuration (1 - 16 processes with up to 4 processes per node where each process checkpoints 8 GB of data; higher is better)}
    \label{fig:synthetic-CRs-writes-multinode}
\endminipage
\hfill
\minipage{0.32\textwidth}
    \centering
    \includegraphics[width=\linewidth]{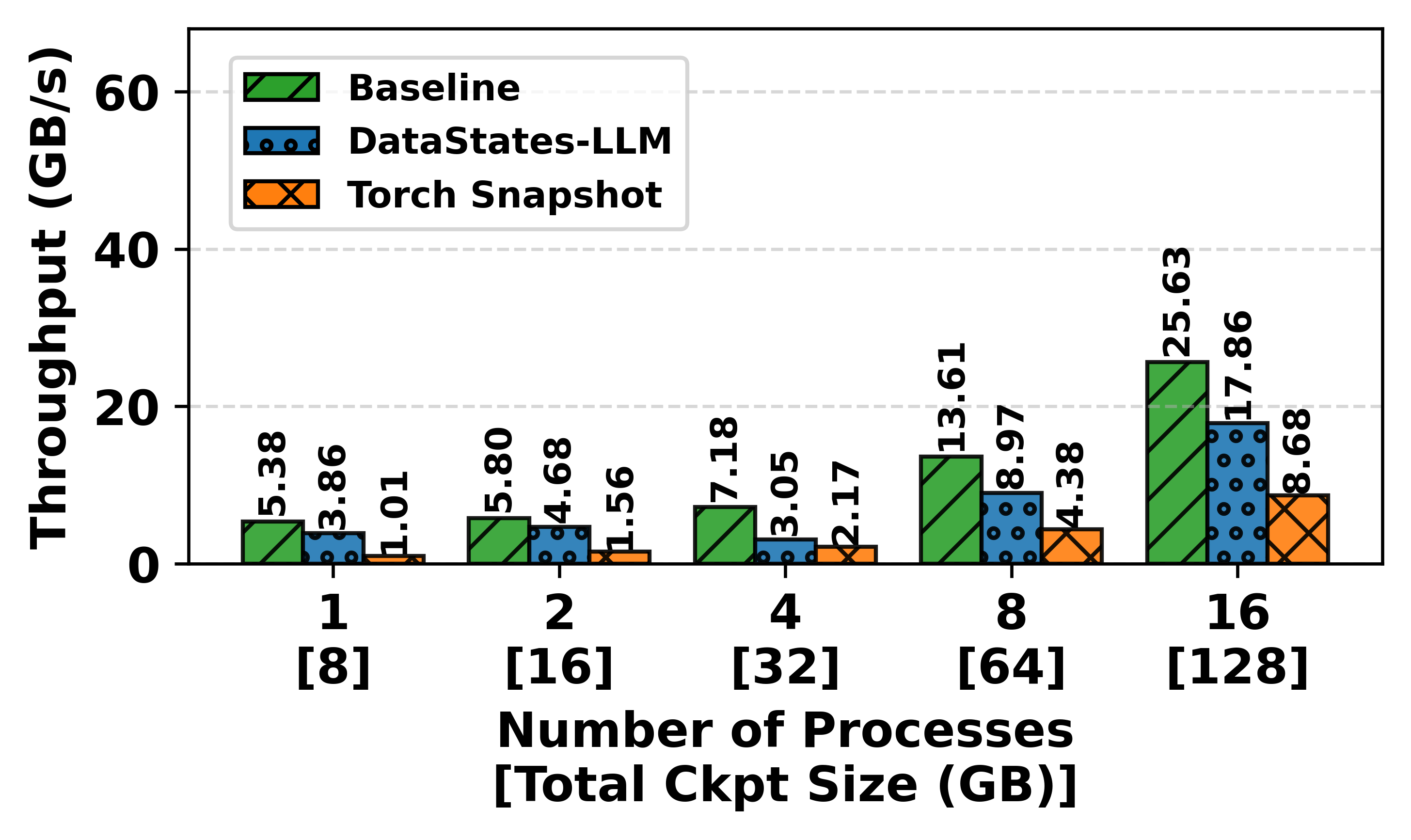}
    \Description{bar chart comparing read performance using different checkpointing engines (DataStates-LLM and TorchSnapshot) to the liburing baseline on the synthetic benchmark. The liburing baseline outperforms both checkpointing engines}
    \caption{Restore throughput of the synthetic benchmark using the single aggregated file configuration (1 - 16 processes with up to 4 processes per node where each process restores 8 GB of data; higher is better)}
    \label{fig:synthetic-CRs-reads-multinode}
\endminipage
\hfill
\minipage{0.32\textwidth}
    \centering
    \includegraphics[width=\linewidth]{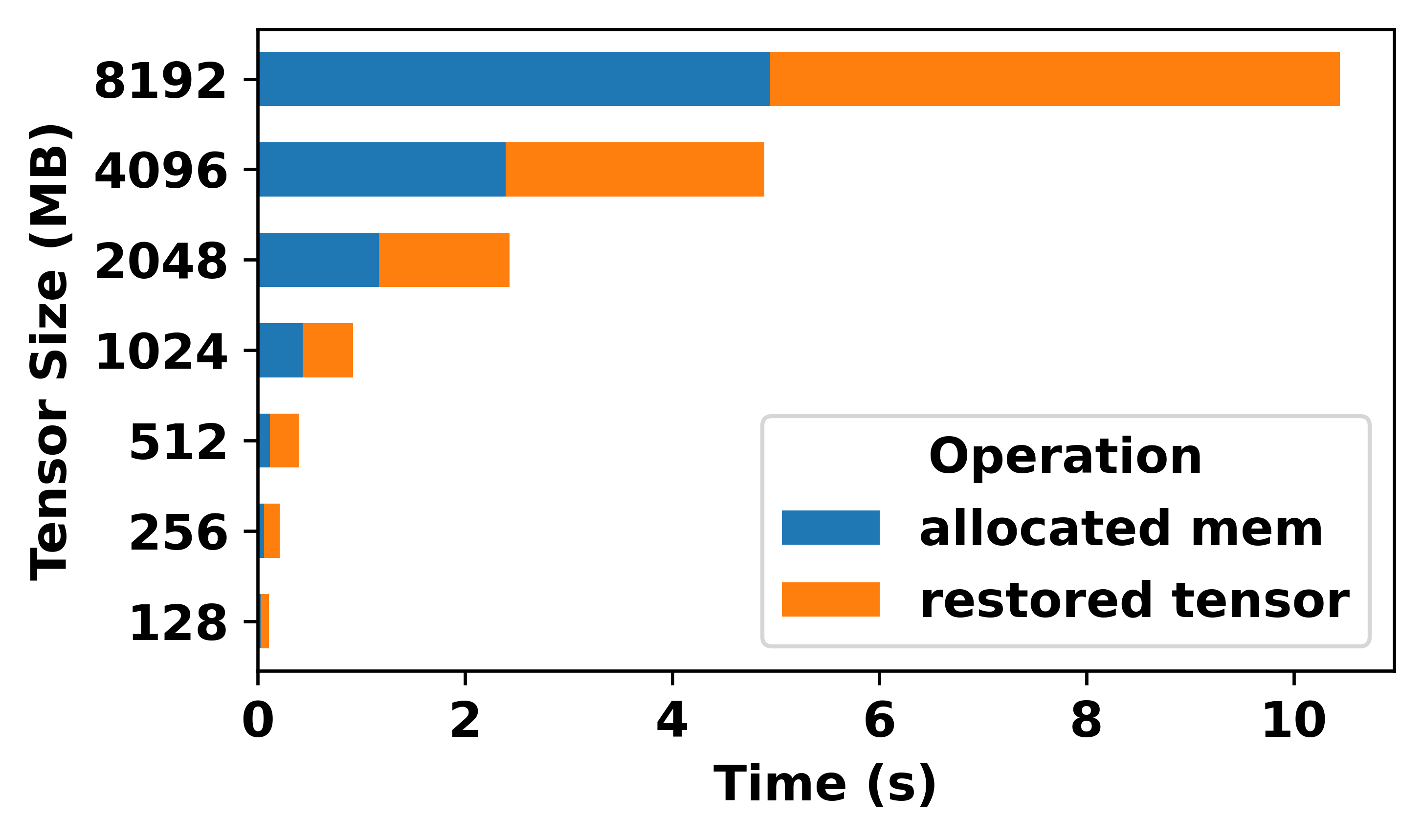}
    \Description{horizontal bar chart showing time in seconds it takes to perform memory allocation and the rest of the restore pipeline for DataStates-LLM highlighting the impact of the memory wall}
    \caption{DataStates-LLM restore pipeline broken down by major operations-- memory allocation and PFS reads (1 compute node with 4 processes)}
    \label{fig:datastates-pipeline-time}
\endminipage
\hfill
\end{figure*}

\begin{figure*}
\minipage{0.32\textwidth}
    \centering
    \includegraphics[width=\linewidth]{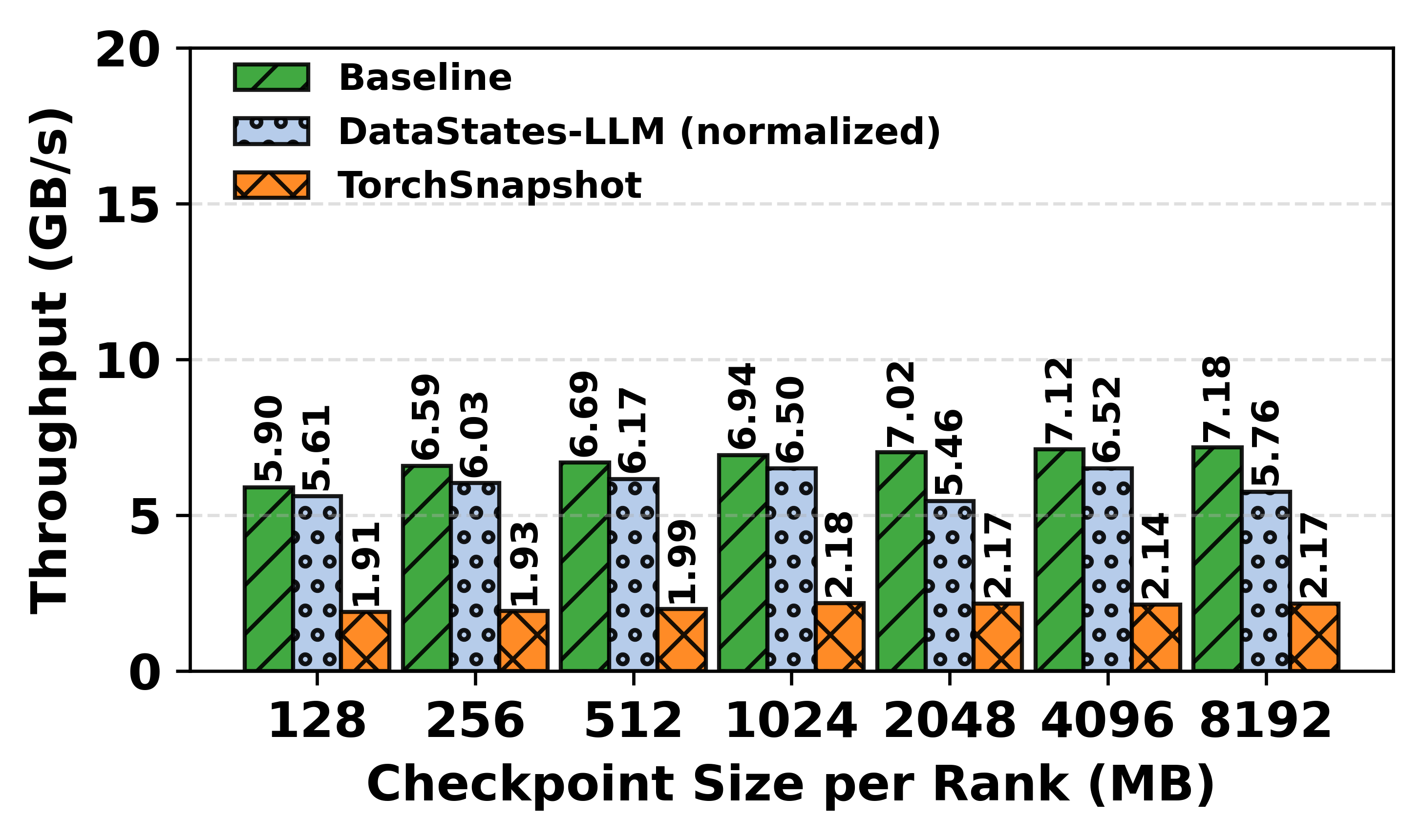}
    \Description{bar chart comparing read performance using different checkpointing engines (normalized DataStates-LLM and TorchSnapshot) to the liburing baseline on the synthetic benchmark varying tensor sizes by powers of 2. The liburing baseline outperforms both checkpointing engines but the normalized DataStates-LLM is much closer to the baseline compared to before}
    \caption{Restore throughput of the synthetic benchmark using the single aggregated file configuration. DataStates-LLM does not include memory allocation time (1 compute node with 4 processes; higher is better)}
    \label{fig:datastates-pipeline-throughput}
\endminipage
\hfill
\minipage{0.32\textwidth}
    \centering
    \includegraphics[width=\linewidth]{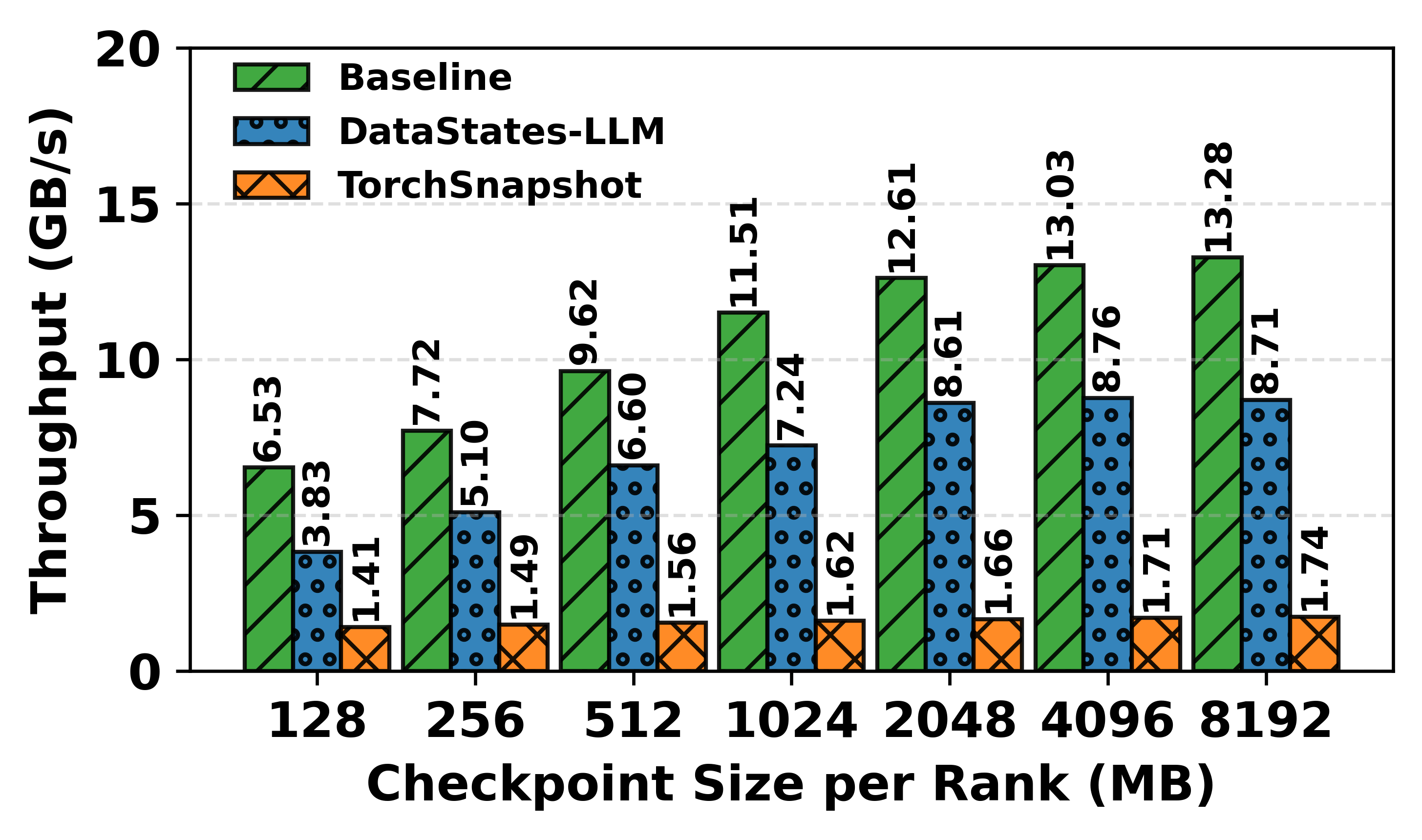}
    \Description{bar chart comparing write performance using different checkpointing engines (DataStates-LLM and TorchSnapshot) to the liburing baseline on the synthetic benchmark varying tensor sizes by powers of 2. The liburing baseline outperforms both checkpointing engines}
    \caption{Checkpoint throughput of the synthetic benchmark using the single aggregated file configuration (1 compute node with 4 processes; higher is better)}
    \label{fig:CR-singlenode-write}
\endminipage
\hfill
\minipage{0.32\textwidth}
    \centering
    \includegraphics[width=\linewidth]{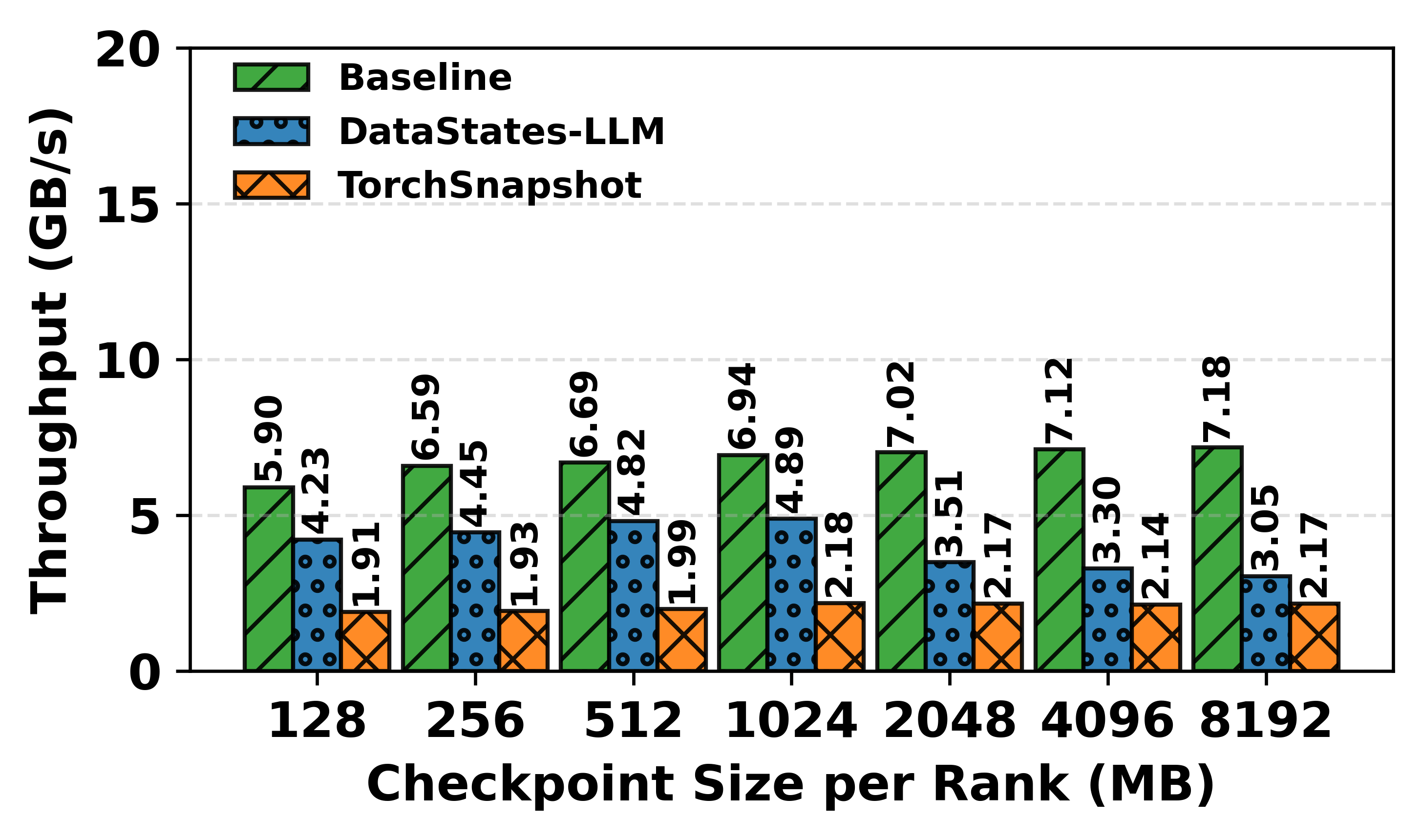}
     \Description{bar chart comparing read performance using different checkpointing engines (DataStates-LLM and TorchSnapshot) to the liburing baseline on the synthetic benchmark varying tensor sizes by powers of 2. The liburing baseline outperforms both checkpointing engines}
    \caption{Restore throughput of the synthetic benchmark using the single aggregated file configuration (1 compute node with 4 processes; higher is better)}
    \label{fig:CR-singlenode-read}
\endminipage
\end{figure*}

\fcolorbox{black}{purple!8}{%
  \begin{minipage}{0.95\linewidth}
\textbf{Observation 2:} The impact of \texttt{O\_DIRECT} on I/O performance diverges between checkpoint and restart phases of LLM training. While direct I/O consistently improves checkpoint write throughput by avoiding double buffering, its effect on restore performance is highly dependent on access granularity, working-set size, and access ordering. Unlike traditional scientific restart workloads, LLMs checkpoints consist of largely non-uniform tesnor layouts ranging from KB or MB to multi-gigabyte parameter shards. Restore phasses therefore interleave small, fragmented reads with large, sequential transfers, limiting page-cache reuse and increaseing management overhead. The assymetry, combined with the increasing frequency of restarts during LLM training and inference workloads, inidicates that caching assumptions inherited from traditional HPC workloads do not hold for large-scale LLM restore patterns and must be explicitly accounted for in checkpoint system design. 
  \end{minipage}
}

\subsection{Modeling Ideal \texttt{liburing} Performance Against LLM Checkpointing Engines}
\label{sec:synthetic-CR-comparisons-results}
We next compare our liburing-based synthetic benchmark (single aggregated file configuration) against two production-grade LLM checkpointing systems: DataStates-LLM and TorchSnapshot. Both represent state-of-the-art frameworks, but differ in backend and design: DataStates-LLM uses liburing, while TorchSnapshot relies on libaio, an older interface lacking modern batching and queueing capabilities. This comparison highlights how different I/O engines, abstractions, and runtime design affect end-to-end C/R throughput.

Figures~\ref{fig:synthetic-CRs-writes-multinode}–\ref{fig:synthetic-CRs-reads-multinode} show that our baseline achieves up to 1.2$\times$ and 6.6$\times$ higher write and 1.5$\times$ and 3$\times$ higher read throughput than DataStates-LLM and TorchSnapshot, respectively. TorchSnapshot’s performance collapses under heavy metadata and fragmentation overhead: each object ($\leq$512 MB) is split into multiple files and directories, amplifying MDS/OSS contention and I/O serialization. Combined with the limitations of libaio, these design choices lead to severe throughput degradation and no scalability.

Although DataStates-LLM uses the same liburing backend as our benchmark, its performance is constrained by higher-level runtime costs. Our benchmark isolates raw I/O throughput, while DataStates-LLM executes a full checkpoint/restore pipeline involving metadata management, object serialization, and dynamic buffer allocation. Thus, its lower throughput reflects framework overheads rather than I/O inefficiency. However, DataStates-LLM’s write throughput plateaus beyond 2~GB per rank, and read throughput declines beyond 1~GB, indicating additional bottlenecks.

To pinpoint these bottlenecks, we profiled the DataStates-LLM restore path. As depicted in Figure~\ref{fig:datastates-pipeline-time}, memory allocation dominates restore time, nearly matching raw read cost, while other operations such as deserialization and header parsing are negligible. Excluding allocation overhead nearly doubles throughput (Figure~\ref{fig:datastates-pipeline-throughput}), aligning it with the baseline. This confirms that restore performance is primarily memory-bound, suggesting that preallocated or circular buffer reuse could significantly improve efficiency.

\vspace{10pt}
\fcolorbox{black}{purple!8}{%
  \begin{minipage}{0.95\linewidth}
{\bf Observation 3:}
Production frameworks underperform relative to the isolated \texttt{liburing} baseline due to design and runtime overheads rather than I/O inefficiency. TorchSnapshot’s fragmentation and metadata-heavy structure severely limit scalability, while DataStates-LLM’s throughput plateaus from per-object allocation overheads. Profiling shows that memory allocation dominates restore latency-- removing it nearly doubles throughput-- indicating that preallocated or reusable buffers could close much of the performance gap.
  \end{minipage}
}

\subsection{Impacts of LLM Checkpointing Structures}

\begin{figure*}[ht]
\minipage{0.49\textwidth}
    \centering
    \includegraphics[width=\linewidth]{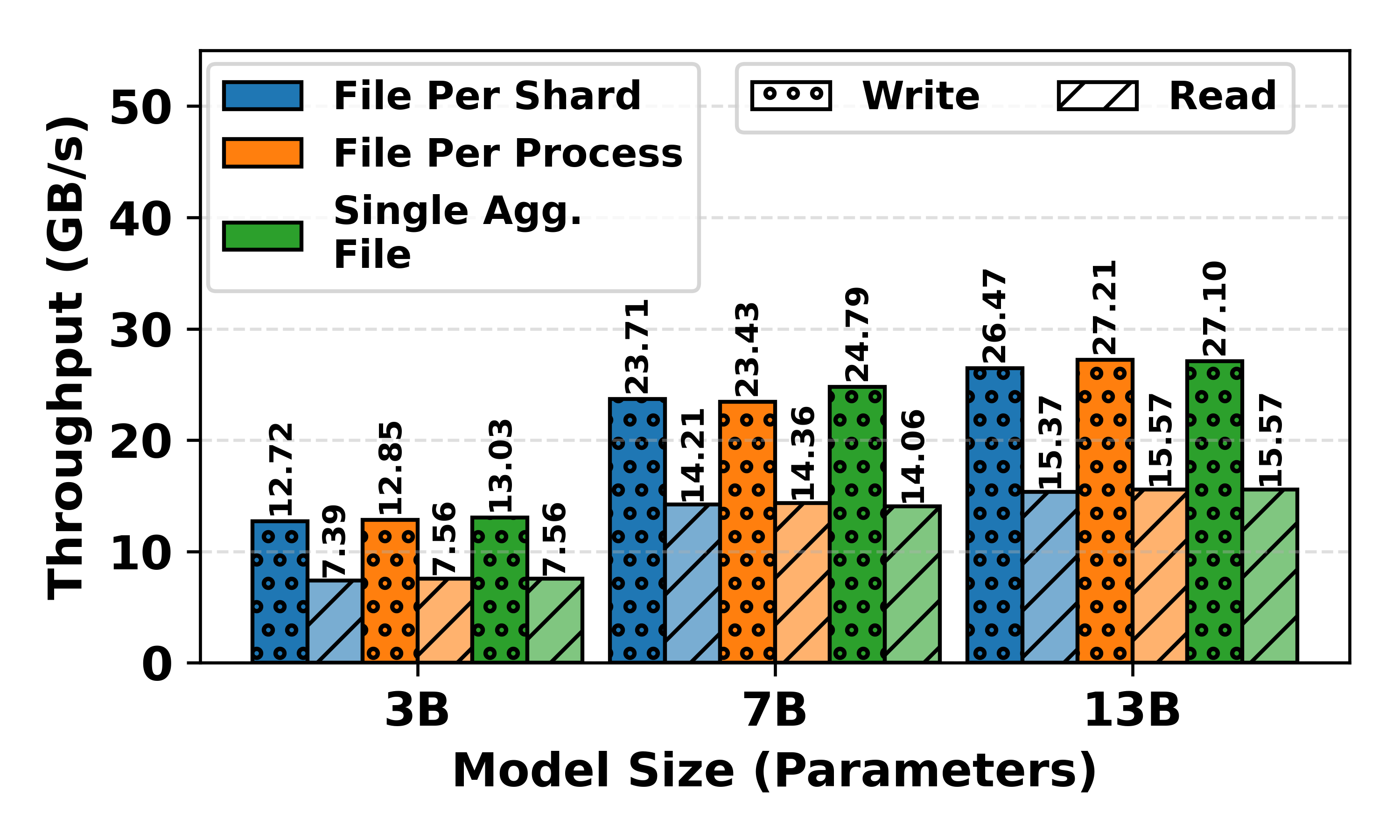}
     \Description{bar chart comparing read and write performance for different aggregation strategies using the representative LLM-Benchmark showing marginal performance difference across aggregation strategies}
    \caption{Read and write throughputs with different aggregation strategies on the realistic LLM benchmark using 3B (4 GPUs), 7B (8 GPUs), and 13B (16 GPUs) models, 4 GPUs/node; higher is better.}
    \label{fig:LLM-agg-scalability}
\endminipage
\hfill
\minipage{0.49\textwidth}
    \centering
    \includegraphics[width=\linewidth]{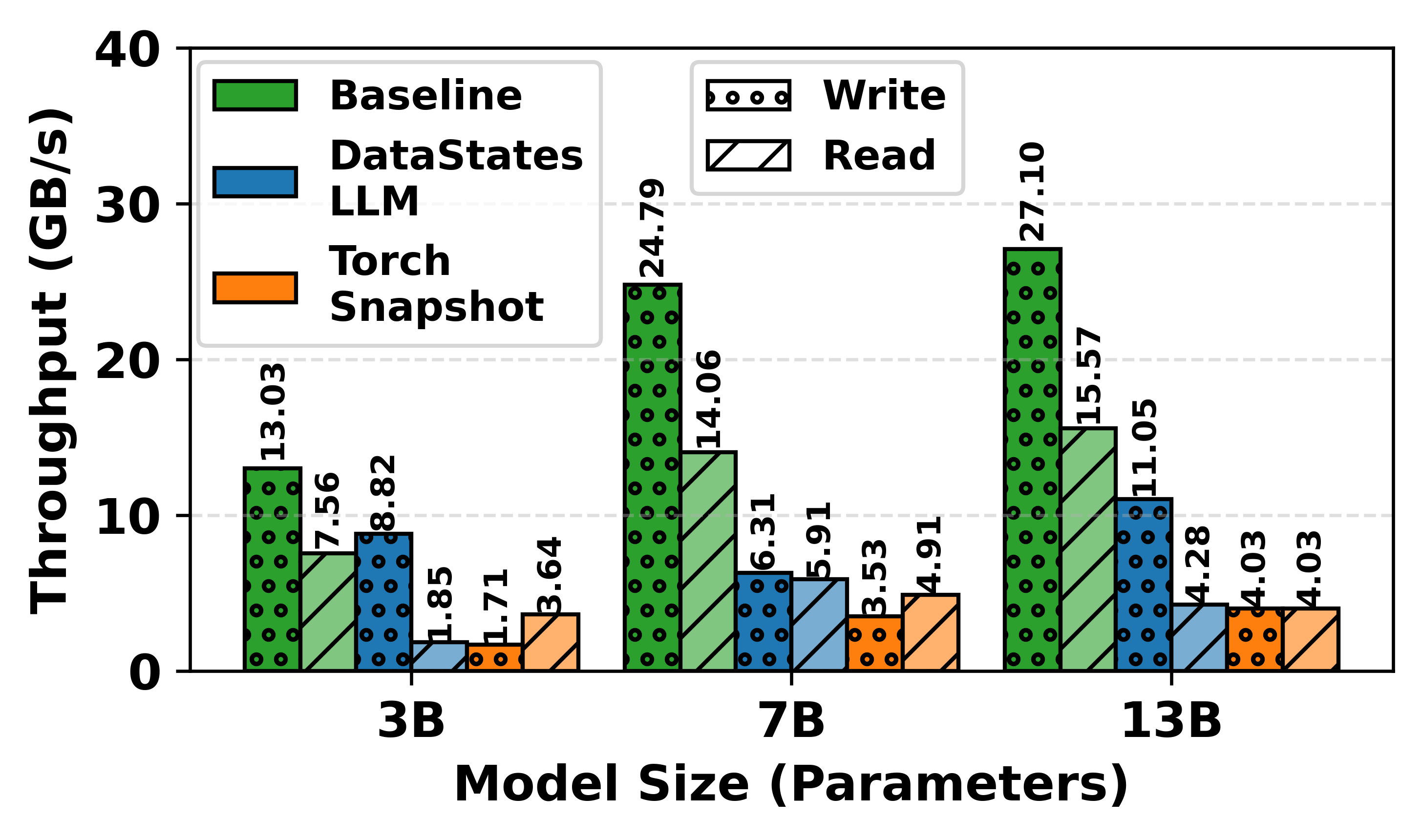}
    \Description{bar chart comparing read and write performance for different checkpointing engines (DataStates-LLM and TorchSnapshot) to the liburing baseline using the representative LLM-Benchmark showing liburing baseline presents promising performance gains for both read and write improvement}
    \caption{Checkpoint and restore throughputs of the realistic LLM benchmark using the single aggregated file configurations;  using 3B (4 GPUs), 7B (8 GPUs), and 13B (16 GPUs) models, 4 GPUs/node; higher is better.}
    \label{fig:full_comparisons}
\endminipage
\end{figure*}

We evaluate our \texttt{liburing} baseline under realistic LLM checkpoint/restore workloads (\S\ref{sec:LLM-bench-design}). While both experiments use the same I/O engine, their configurations differ fundamentally: the synthetic benchmark uses a single large contiguous buffer per process, whereas the LLM benchmark reflects true model layouts—multiple buffers of varying, non-aligned sizes.
Unlike the synthetic setup, which used block-aligned power-of-two regions, the LLM benchmark requires explicit offset alignment for each buffer. Under the single aggregated file configuration, this introduces a serialized prefix-sum calculation across ranks: each must wait for its predecessor to compute total space (including padding) before writing, slightly increasing coordination overhead. These layout constraints stress different parts of the I/O stack: the synthetic benchmark primarily measures raw data-path throughput, while the LLM benchmark exposes metadata management, alignment handling, and queue depth limits that better resemble end-to-end checkpoint behavior. In particular, the irregular write sizes reduce batching opportunities within \texttt{liburing}, often resulting in shorter submission queues and lower completion parallelism

Figure~\ref{fig:LLM-agg-scalability} shows throughput for 3B, 7B, and 13B checkpoints across aggregation strategies. Unlike the synthetic experiments (Figures~\ref{fig:synthetic_aggregation_multinode_write}–\ref{fig:synthetic-agg-singlenode-reads}), where aggregation yielded clear gains over file-per-shard, all strategies here perform comparably, with only modest aggregation benefits.

This difference stems from non-coalesced I/O. In the synthetic case, each rank could submit one large batched request, maximizing queue concurrency. In contrast, the LLM benchmark issues many small, irregularly sized buffers. The effect intensifies for larger checkpoints (7B, 13B), where a larger percentage of I/O operations are using small sizes, halving sustained throughput compared to the synthetic baseline. For example, 13B contains many small ($\leq$5 MB) buffers, each generating separate I/O requests that limit coalescing and increase metadata activity.

Despite this, aggregated layouts yield tangible throughput gains and greatly simplifies checkpoint management. These findings suggest that future checkpointing frameworks could benefit from hybrid aggregation strategies (e.g., combining logical object grouping in user space with kernel) level request batching—to exploit \texttt{liburing}’s concurrency more effectively.

When comparing checkpointing frameworks in Figure~\ref{fig:full_comparisons} using the best-performing aggregation strategy (single aggregated file), our benchmark achieves consistently higher throughput than both DataStates-LLM and TorchSnapshot, showing larger relative gains under realistic LLM workloads than in the synthetic benchmark.

Across all three model sizes, our benchmark achieves up to $3.9\times$ higher write throughput than DataStates-LLM and $7.6\times$ higher than TorchSnapshot. For reads we observe up to $3.6\times$ improvement over DataStates-LLM and $3.8\times$ over TorchSnapshot. These gains are particularly pronounced for larger models, where there are more numbers of small files and buffers ($\leq 5.1$ MB) increase metadata overhead and utilize available bandwidth less efficiently by submitting many small I/O requests.
Despite its simplicity, this design achieves superior end-to-end throughput in both synthetic and LLM-replica benchmarks, highlighting that both I/O request aggregation, as well as file aggregation offers a promising avenue for scalable, high-performance C/R frameworks for next generation AI workloads.

\vspace{10pt}
\fcolorbox{black}{purple!8}{%
  \begin{minipage}{0.95\linewidth}
    {\bf Observation 4:}
    Realistic LLM checkpoint layouts expose the performance penalties of uncoalesced I/O, where fragmented and variably sized buffers disrupt batching and offset alignment. While aggregation still improves throughput over the default file-per-shard approach, small, misaligned objects and serialized offset computations limit achievable concurrency and reduce sustained bandwidth. Compared to production frameworks, the streamlined \texttt{liburing} baseline sustains over 3$\times$ higher write and read throughput, showing that efficient batching and object aggregation are key to scalable C/R under real workloads.
  \end{minipage}
}

\section{Discussion}

Our experiments highlight the critical impact of aggregation and the O\_DIRECT flag on LLM checkpoint/restore I/O.
Our results clearly show that using a single aggregated file layout consistently outperforms the default file-per-shard approach adopted by current C/R engines, as well as the more aggressively sharded layouts used by TorchSnapshot, showing we can obtain better performance and significantly improve checkpoint management for users. Furthermore, by coalescing multiple small objects (e.g., layer weights, model states, and metadata) into larger I/O operations, we show that performance can improve bandwidth utilization and reduce metadata contention.

The impact of the O\_DIRECT flag is strongly workload dependent. For writes, enabling O\_DIRECT bypasses page and OSS cahces, improving throughput by up to $4.8\times$. For reads of small objects or files, O\_DIRECT is shown to reduce performance, as it cannot take advantage of cached data that reduces PFS interactions and improves throughput. However, for larger read operations ($\geq4$GB), it improves read throughput by avoiding cache coherency overhead in both the kernel and OSS, demonstrating that careful use of O\_DIRECT can optimize performance depending on read/write size and workload characteristics (e.g, write-heavy vs. read-heavy).

Comparing against TorchSnapshot and DataStates-LLM highlights the remaining gap to peak achievable performance. Compared to our aggregated baseline approach, both C/R engines fall short of of synthetic benchmark peaks, trailing by up to $1.2\times$ in writes and $1.5\times$ in reads for DataStates-LLM and up to $6.6\times$ in writes and $3\times$ in reads for TorchSnapshot. These differences are even more pronounced in the LLM benchmark, where DataStates-LLM trails the baseline benchmark by up tp $3.9\times$ for writes, and $3.6\times$ for reads. Comparatively, TorchSnapshot trails by up to $6.7\times$ in write throughput, and $3.8\times$ in read performance. A key limitation in DataStates-LLM is repeated buffer allocation during reads, whereas our baseline implementation reuses preallocated, aligned buffers, reducing memory overhead and improving read throughput. These observations indicate significant room for improvement via aggregation-aware I/O, coalescing of small I/O operations, and careful alignment, which can help future frameworks more fully exploit available bandwidth and minimize metadata bottlenecks.

\section{Conclusions and Future Work}
Our study demonstrates that careful aggregation of checkpoint data can substantially improve both write and read throughput in LLM training. By consolidating data (such as layer weights, optimizer states, and metadata) into single, contiguous files, we reduce metadata overhead, improve bandwidth utilization, and simplify restore operations by limiting each process to a single file handle. We further show that the O\_DIRECT flag has a workload-dependent impact: it greatly accelerates writes by bypassing page and OSS caches, while for small reads it can reduce performance; for large reads ($\leq4$GB), however, it mitigates cache coherency overhead, improving throughput. These results highlight that combining aggregation-aware I/O with judicious use of O\_DIRECT can significantly enhance checkpoint/restore efficiency

Future work will focus on integrating these insights into production frameworks such as DataStates-LLM. Specifically, we plan to implement persistent buffer reuse to reduce repeated memory allocations, batched I/O submissions aligned with queue depth, and multi-threaded completion handling to fully utilize available bandwidth. In addition, we aim to coalesce small objects (e.g., layer weights, optimizer states, metadata) into larger I/O operations to maximize bandwidth utilization and reduce metadata pressure. By incorporating these optimizations, next-generation C/R systems can achieve scalable, high-performance checkpointing tailored to the demands of large LLM training pipelines.

\begin{acks}
This work is supported in part by the U.S. Department of Energy (DOE), Office of Advanced Scientific Computing Research (ASCR) under contract DEAC02--06CH11357/0F--60169 and the National Science Foundation (NSF) under award no.\  2411386/2411387, 2106635. Results presented in this paper are obtained using Argonne's HPC systems-- ALCF Polaris, Joint Laboratory for System Evaluation (JLSE), and NSF's CloudLab and Chameleon testbeds.
\end{acks}

\balance
\bibliographystyle{ACM-Reference-Format}
\bibliography{main}

\end{document}